\documentclass{LMCS}

\def\doi{8 (1:16) 2012}
\lmcsheading%
{\doi}
{1--42}
{}
{}
{May~\phantom.16, 2011}
{Mar.~\phantom02, 2012}
{}

\usepackage{enumerate,hyperref,hareng}

\hypersetup{%
pdftitle={Deciding Kleene Algebras in Coq},%
pdfauthor={Thomas Braibant, Damien Pous},%
pdfkeywords={Coq proof assistant, reflexive tactic, Kleene algebra,
  finite automata, regular expressions, decision procedure, typeclasses}%
}

\begin{document}

\title[Deciding Kleene Algebras in Coq]{Deciding Kleene Algebras in Coq\rsuper*}

\author{Thomas Braibant}
\author{Damien Pous}
\address{CNRS, INRIA, LIG, UMR 5217, Université de Grenoble, France, Europe}
\email{\{thomas.braibant,damien.pous\}@inria.fr}
\thanks{This work was supported by the project PiCoq (ANR 2010 BLAN 0305 01).}

\titlecomment{{\lsuper*}This paper is a long version of the abstracts we
  presented at the first Coq workshop (August 2009) and in Proc. 1st
  ITP, vol. 6172 of LNCS, 2010~\cite{BraibantP10}.}

\begin{abstract}
  \noindent 
  We present a reflexive tactic for deciding the equational theory of
  Kleene algebras in the Coq proof assistant. This tactic relies on a
  careful implementation of efficient finite automata algorithms, so
  that it solves casual equations instantaneously and properly scales
  to larger expressions. The decision procedure is proved correct and
  complete: correctness is established w.r.t. any model by formalising
  Kozen's initiality theorem; a counter-example is returned when the
  given equation does not hold. The correctness proof is challenging:
  it involves both a precise analysis of the underlying automata
  algorithms and a lot of algebraic reasoning. In particular, we have
  to formalise the theory of matrices over a Kleene algebra. We build
  on the recent addition of first-class typeclasses in Coq in order to
  work efficiently with the involved algebraic structures.
\end{abstract}

\maketitle

\section{Introduction}
\label{sec:intro}

\subsection{Motivations.}
\label{ssec:motiv}

Proof \emph{assistants} like Coq or Isabelle/HOL make it possible to
leave technical or administrative details to the computer, by defining
high-level tactics. For example, one can define tactics to solve
decidable problems automatically (e.g., \code{omega} for Presburger
arithmetic and \code{ring} for ring equalities). Here we present a
tactic for solving equations and inequations in Kleene algebras. This
tactic belongs to a larger project whose aim is to provide tools for
working with binary relations in Coq. Indeed, Kleene algebras
correspond to a non-trivial decidable fragment of binary relations. In
the long term, we plan to use these tools to formalise results in
rewriting theory, process algebras, and concurrency theory results.
Binary relations play a central role in the corresponding semantics.

\begin{figure}[t]
  \centering
  \begin{minipage}{.30\linewidth}
    \begin{align*}
      \inferrule{%
        \xymatrix@R=.6em {%
          &\cdot\ar[rd]^{S^\star} \ar@{}[dd]|H\\
          \cdot\ar[ru]^R\ar@{.>}[dr]_{S^\star} &&\cdot\\
          &\cdot\ar@{.>}[ur]_{R^\star}}\\}{%
        \xymatrix@R=.6em {%
          &\cdot\ar[rd]^{S^\star} &&\cdot\\
          \cdot\ar[ru]^R\ar@{.>}[dr]_{S^\star} && \cdot\ar[ru]^{R^\star}\\
          &{?}\ar@{.>}@/_1.5em/[uurr]_>(.7){R^\star}& }}
    \end{align*}
  \end{minipage}\hfill
  \begin{minipage}{.34\linewidth}
    \begin{coq}
R,S: relation P
H: forall p,r,q, R p r -> S# r q 
    -> exists s, S# p s /\ R# s q
p,q,q',s: P 
Hpq: R p q 
Hqs: S# q s 
Hsq': R# s q'
\hline
exists s, S# p s /\ R# s q'
    \end{coq}    
  \end{minipage}\hfill
  \begin{minipage}{.2\linewidth}
    \begin{coq}





R,S: X
H: R@S# <== S#@R#
\hline
R@S#@R# <== S#@R#
    \end{coq}    
  \end{minipage}
  \caption{Diagrammatic, concrete, and abstract presentations of the
    same state in a proof.}
  \label{fig:cr}
\end{figure}

A starting point for this work is the following remark: proofs about
abstract rewriting (e.g.,\ Newman's Lemma, equivalence between weak
confluence and the Church-Rosser property, termination theorems based
on commutation properties) are best presented using informal ``diagram
chasing arguments''. This is illustrated by Fig.~\ref{fig:cr}, where
the same state of a typical proof is represented three times. Informal
diagrams are drawn on the left. The goal listed in the middle
corresponds to a naive formalisation where the points related by
relations are mentioned explicitly. This is not satisfactory: a lot of
variables have to be introduced, the goal is displayed in a rather
verbose way, the user has to draw the intuitive diagrams on its own
paper sheet. On the contrary, if we move to an algebraic setting (the
right-hand side goal), where binary relations are seen as abstract
objects that can be composed using various operators (e.g.,\ union,
intersection, relational composition, iteration), statements and Coq's
output become rather compact, making the current goal easier to read
and to reason about.

More importantly, moving to such an abstract setting allows us to
implement several decision procedures that could hardly be stated with
the concrete presentation.  For example, after the user rewrites the
hypothesis \code H in the right-hand side goal of Fig.~\ref{fig:cr},
we obtain the inclusion \coqe{S#@R#@R# <== S#@R#}, which is a
(straightforward) theorem of Kleene algebras: the tactic we describe
in this paper proves this sub-goal automatically.

\subsection{Mathematical background}
\label{ssec:deka}

A Kleene algebra~\cite{Koz94b} is a tuple
$\tuple{X,\cdot,+,1,0,\star}$, where $\tuple{X,\cdot,+,1,0}$ is an
idempotent non-commutative semiring, and $\_^\star$ is a unary post-fix
operation on $X$, satisfying the following axiom and inference rules
(where $\le$ is the partial order defined by $x\le y ~ \triangleq ~
x+y = y)$:
\begin{mathpar}
  1+x\cdot x^\star = x^\star \and
  \inferrule{x\cdot y \le y}{x^\star\cdot y \le y} \and %
  \inferrule{y\cdot x \le y}{y\cdot x^\star \le y}
\end{mathpar}
Terms of Kleene algebras, ranged over using $x,y$, are called
\emph{regular expressions}, irrespective of the considered model.
Models of Kleene algebras include \emph{languages}, where the unit
$(1)$ is the language reduced to the empty word, product $(\cdot)$ is
language concatenation, and star $(\_^\star)$ is language iteration;
and \emph{binary relations}, where the unit is the identity relation,
product is relational composition, and star is reflexive and
transitive closure. Here are some theorems of Kleene algebras:
\begin{mathpar}
  x^\star = x^\star\cdot x^\star = {x^\star}^\star = (x+1)^\star \and 
  (x+y)^\star = x^\star\cdot{(y\cdot x^\star)}^\star \and
  x\cdot(y\cdot x)^\star = (x\cdot y)^\star\cdot x 
\end{mathpar}

\noindent %
Among languages, those that can be described by a
finite state automaton (or equivalently, generated by a regular
expression) are called \emph{regular}.  Thanks to finite automata
theory~\cite{Kle56,RabinScott}, equality of regular languages is
decidable:
\begin{quote}\em
  ``two regular languages are equal if and only if the corresponding
  minimal automata are isomorphic''.
\end{quote}
However, the above theorem is not sufficient to derive equations in
all Kleene algebras: it only applies to the model of regular
languages. We actually need a more recent theorem, by
Kozen~\cite{Koz94b} (independently proved by Krob~\cite{Krob91a}):
\begin{quote}\em 
  ``if two regular expressions $x$ and $y$ denote the same regular
  language, then $x=y$ is a theorem of Kleene algebras''.
\end{quote}
In other words, the algebra of regular languages is initial among
Kleene algebras: we can use finite automata algorithms to solve
equations in an arbitrary Kleene algebra.

The main idea of Kozen's proof is to encode finite automata using
matrices over regular expressions, and to replay the algorithms at
this algebraic level. Indeed, a finite automaton can be represented
with three matrices $\tuple{u,M,v}\in \mathcal M_{1,n}\times\mathcal
M_{n,n}\times\mathcal M_{n,1}$: $n$ is the number of states of the
automaton, $u$ and $v$ are 0-1 vectors respectively coding for the
sets of initial and accepting states, and $M$ is the transition
matrix: $M_{i,j}$ labels transitions from state $i$ to state $j$.
Consider for example the following non-deterministic automaton, with
three states (like for the automata to be depicted in the sequel,
accepting states are marked with two circles, and short, unlabelled
arrows point to the starting states):
\begin{align*}
  \xymatrix @C=.9em {\ar[r]&
   *++[o][F] {0}\ar@/^.4em/[rr]^a\ar@/^1.7em/[rrrr]^b && 
   *++[o][F=]{1}\ar@/^.4em/[ll]^c\ar[rr]^c &&
   *++[o][F=]{2}\ar@(dr,ur)_{a,b}}
\end{align*}
This automaton can be represented using the following matrices:
\begin{align*}
  u&=\left[%
    \begin{array}{ccc}
      1&0&0\\
    \end{array}
  \right]&%
 M&=\left[%
    \begin{array}{ccc}
      0&a&b\\
      c&0&c\\
      0&~0~&a+b\\
  \end{array}
  \right]&%
  v&=\left[%
    \begin{array}{c}
      0\\1\\1\\
    \end{array}
  \right]\enspace.
\end{align*}
We can remark that the product $u\cdot M\cdot v$ is a scalar (i.e.,\ a
regular expression), which can be thought of as the set of one-letter
words accepted by the automaton---in the example, $a+b$. Similarly,
$u\cdot M^2\cdot v=u\cdot M\cdot M\cdot v$ corresponds to the set of
two-letters words accepted by the automaton---here, $a\cdot c+b\cdot
a+b\cdot b$. Therefore, to mimic the behaviour of a finite automaton
and get the whole language it accepts, we just need to iterate over
the matrix $M$. This is possible thanks to another theorem, which
actually is the crux of the initiality theorem: {\em ``matrices over a
  Kleene algebra form a Kleene algebra''}. We hence have a star
operation on matrices, and we can interpret an automaton
algebraically, by considering the product $u\cdot M^\star\cdot
v$. (Again, in the example, we could check that this computation
reduces into a regular expression which is equivalent to $(a\cdot
c)^\star\cdot(a+(b+a\cdot c)\cdot(a+b)^\star)$, which corresponds
precisely to the language accepted by the automaton.)

\subsection{Overview of our strategy.}
\label{ssec:strat}

We define a \emph{reflexive} tactic. This methodology is quite
standard~\cite{BoyerMoore81,DBLP:conf/lics/AllenCHA90}. For example,
this is how the \code{ring} tactic is
implemented~\cite{DBLP:conf/tphol/GregoireM05}. Concretely, this means
that we program the decision procedure as a Coq function, and that we
prove its correctness and its completeness within the proof assistant:
\begin{coq}
Definition decide_kleene: regex -> regex -> bool := ...
Theorem Kozen94: forall x y: regex, decide_kleene x y = true <-> x == y.
\end{coq}

The above statement corresponds to correctness and completeness with
respect to the syntactic ``free'' Kleene algebra: \regex\ is the
inductive type for regular expressions over a countable set of variables,
and \coqe{==} is the inductive equality generated by the axioms of
Kleene algebras and the rules of equational reasoning. 
Using reification mechanisms, this is sufficient for our needs: the
result can be lifted to other models using simple tactics.

\medskip\noindent %
Here are the main requirements we had to take into account for the
design of the library:

\begin{describe}
\item[Efficiency] The equational theory of Kleene algebras is
  PSPACE-complete~\cite{MS73}; this means that the \dk{} function must
  be written with care, using efficient algorithms.
  Notably, the matricial representation of automata is not efficient,
  so that formalising Kozen's ``mathematical'' proof~\cite{Koz94b} in
  a naive way would be computationally impracticable. Instead, we need
  to choose appropriate data structures for automata and algorithms,
  and to rely on the matricial representation only in proofs, using
  the adequate translation functions.


\item[Heterogeneous models] Homogeneous binary relations are a model
  of Kleene algebras, but binary relations can be heterogeneous: their
  domain might differ from their co-domain so that they fall out of
  the scope of standard Kleene algebra. We could use a trick to handle
  the special case of heterogeneous relations~\cite{Krauss}, but there
  is a more general and more algebraic solution that captures all
  heterogeneous models: it suffices to consider the rather natural
  notion of \emph{typed} Kleene algebra~\cite{Koz98b}. Since we want
  to put forward the algebraic approach, we tend to prefer this second
  option. Moreover, as pointed out in next paragraph, we actually
  exploit this generalisation to formalise Kozen's proof.

\item[Matrices] As explained in Sect.~\ref{ssec:deka}, Kozen's proof
  relies on the theory of matrices over regular expressions, which we
  thus need to formalise. First, this formalisation must be tractable
  from the proof point of view: the overall proof requires a lot of
  matricial reasoning. Second, we must handle rectangular matrices,
  which appear in some parts of the proof (see
  Sect.~\ref{ssec:determ}). The latter point can be achieved in a nice
  way thanks to the generalisation to typed Kleene algebra: while only
  square matrices of form a model of Kleene algebra, rectangular
  matrices form a model of typed Kleene algebra.

\item[Sharing] The overall proof being rather involved, we need to
  exploit sharing as much as possible. For instance, we work with
  several models of Kleene algebra (the syntactic model of regular
  expressions, matrices over regular expressions, languages, matrices
  over languages, and relations). Since these models share the same
  properties, we need to share notation, basic laws, theorems, and
  tactics: this improves readability, usability, and
  maintainability. Similarly, the proof requires vectors, which we
  define as a special case of (rectangular) matrices: this saves us
  from re-developing their theory separately.

\item[Modularity] Following mathematical and programming practice, we
  aim at a modular development: this is required to be able to get
  sharing between the various parts of the proof. A typical example is
  the definition of the Kleene algebra of matrices
  (Sect.~\ref{ssec:mxstar}), which corresponds to a rather long
  proof. With a monolithic definition of Kleene algebra, we would have
  to prove that all axioms of Kleene algebra hold \emph{from
    scratch}. On the contrary, with a modular definition, we can first
  prove that matrices form an idempotent semiring, which allows us to
  use theorems and tactics about semiring when proving that the
  defined star operation actually satisfies the appropriate laws.

\item[Reification] The final tactic (for deciding Kleene algebras) and
  some intermediate tactics are defined by reflection. Therefore, we
  need a way to achieve \emph{reification}, i.e., to transform a goal
  into a reified version that lets us perform computations within
  Coq. Since we work with \emph{typed} models, this step is more
  involved than is usually the case.

\end{describe}

\subsection*{Outline of the paper.}
Section~\ref{sec:underlying} is devoted to the underlying design
choices. We explain how we define matrices in Sect.~\ref{sec:mx}. The
algorithm and its correctness proof are described in
Sect.~\ref{sec:algo}. We discuss the efficiency of the tactic in
Sect.~\ref{sec:perfs}. We conclude with related works and directions
for future work in Sect.~\ref{sec:ccl}.

\section{Underlying design choices}
\label{sec:underlying}

According to the above constraints and objectives, an essential
decision was to build on the recent introduction of first-class
\emph{typeclasses} in Coq~\cite{SozeauOury08}. This section is devoted
to the explanation of our methodology: how to use typeclasses to
define the algebraic hierarchy in a modular way, how to formalise
typed algebras, how to reify the corresponding expressions.  We start
with a brief description of the implementation of typeclasses in Coq.

\subsection{Basic introduction to typeclasses in Coq.}
\label{ssec:intro:tc}

The overall behaviour of Coq typeclasses~\cite{SozeauOury08} is quite
intuitive; here is how we would translate to Coq a simple Haskell
program that exploits a typeclass \code{Hash} to get a number out of
certain kind of values:

\medskip
\begin{twolistings}
\begin{haskell}
class Hash a where
 hash :: a -> Int
instance Hash Int where
 hash = id
instance (Hash a) => (Hash [a]) where
 hash = sum . map hash
main = print 
 (hash 4, hash [4,5,6], hash [[4,5],[]])
\end{haskell}
&
\begin{coq}
Class Hash A := 
 { hash: A -> nat }.
Instance hash_n: Hash nat := 
 { hash x := x }.
Instance hash_l A: Hash A -> Hash (list A) :=
 { hash l := fold_left (fun a x => a + hash x) l 0 }.
Eval simpl in 
 (hash 4, hash [4;5;6], hash [[4;5];[]]).
\end{coq}
\end{twolistings}
\medskip

\noindent%
Coq typeclasses are first-class; everything is done with plain Coq
terms. In particular, the \coqe{Class} keyword produces a record type
(here, a parametrised one) and the \coqe{Instance} keyword acts like a
standard definition. With the above code we get values of the
following types:
\begin{coq}
Hash: Type -> Type                       hash_n: Hash nat
hash: forall A, Hash A -> A -> nat        hash_l: forall A, Hash A -> Hash (list A)
\end{coq}
The function \code{hash} is a \emph{class projection}: it gives access
to a field of the class. The subtlety is that the first two arguments
of this function are implicit: they are automatically inserted by
unification and typeclass resolution. More precisely, when we write
``\coqe{hash [4;5;6]}\!'', Coq actually reads %
``\coqe{@@hash _ _ [4;5;6]}\!'' (the `@name' syntax can be used in Coq
to give all arguments explicitly).  By unification, the first
placeholder has to be \coqe{list nat}, and Coq needs to guess a term
of type \coqe{Hash (list nat)} to fill the second placeholder. This
term is obtained by a simple proof search, using the two available
instances for the class \code{Hash}, which yields %
``\coqe{@@hash_l nat hash_n}''.
Accordingly, we get the following explicit terms for the three calls
to \coqe{hash} in the above example.
\begin{center}
  \begin{tabular}[t]{|l|l|}
    \hline
    input term & explicit, instantiated, term\\\hline
    \mbox{\coqe{hash 4}} & \mbox{\coqe{@@hash nat hash_n 4}} \\
    \mbox{\coqe{hash [4;5;6]}} & \mbox{\coqe{@@hash (list nat) (@@hash_l nat hash_n) [4;5;6]}} \\
    \mbox{\coqe{hash [[4;5];[]]}} & 
    \mbox{\coqe{@@hash (list (list nat)) (@@hash_l (list nat) (hash_l nat hash_n)) [[4;5];[]]}} \\
    \hline
  \end{tabular}
\end{center}

\medskip %
In summary, typeclasses provide overloading (we can use the
\code{hash} function on several types) and allow one to write much
shorter and readable terms, by letting Coq infer the obvious
boilerplate. This concludes our very short introduction to typeclasses
in Coq; we invite the reader to consult~\cite{SozeauOury08} for more
details.

\subsection{Using typeclasses to structure the development.}
\label{ssec:tc}


We use typeclasses to achieve two tasks: 1) sharing and overloading
notation, basic laws, and theorems; 2) getting a modular definition of
Kleene algebra, by mimicking the standard mathematical hierarchy: a
Kleene algebra contains an idempotent semiring, which is itself
composed of a monoid and a semi-lattice. This very small hierarchy is
summarised below.

\medskip
\begin{center}
  \newcommand\st{<:}
  \small\tt
  \begin{tabular}{cr}
    SemiLattice & $\st$\\
    Monoid & $\st$\\
  \end{tabular}
  IdemSemiRing $\st$ KleeneAlgebra 
\end{center}
\medskip

Before we give concrete Coq definitions, recall that we actually want
to work with the \emph{typed} versions of the above algebraic
structures, to be able to handle both heterogeneous binary relations
and rectangular matrices.  The intuition for moving from untyped
structures to typed structures is given in
Fig.~\ref{fig:typed:algebras}: a typical signature for Kleene algebras
is presented on the left-hand side; we need to move to the signature
on the right-hand side, where a set \code T of indices (or types) is
used to restrain the domain of the various operations. These indices
can be thought of as matrix dimensions; we actually moved to a
categorical setting: \code T is a set of objects, \code{X n m} is the
set of morphisms from \code n to \code m, \code{one} is the set of
identities, and \code{dot} is composition. The semi-lattice operations
(\code{plus} and \code{zero}) operate on fixed homsets; Kleene star
operates only on \emph{square} morphisms---those whose source and
target coincide.
\begin{figure}[tb]
\centering
\begin{tabular}{r@{\qquad}|@{\qquad}l}
  \begin{coq}

X: Type.

dot:  X -> X -> X.  
one:  X.
plus: X -> X -> X.  
zero: X.
star: X -> X.

dot_neutral_left: 
 forall x, dot one x = x.
...
  \end{coq}&
  \begin{coq}
T: Type.
X: T -> T -> Type.

dot:  forall n m p, X n m -> X m p -> X n p.
one:  forall n,     X n n.
plus: forall n m,   X n m -> X n m -> X n m.
zero: forall n m,   X n m.
star: forall n,     X n n -> X n n.

dot_neutral_left: 
 forall n m (x: X n m), dot one x = x.
...
  \end{coq}
\end{tabular}
\caption{From Kleene algebras to typed Kleene algebras.}
\label{fig:typed:algebras}
\end{figure}
%

\subsubsection*{Classes for algebraic operations.}

We now can define the Coq classes on which we based our library. We
first define three classes, for the operations corresponding to a
monoid, a semi-lattice, and Kleene star. These classes are given in
Fig.~\ref{fig:classes:ops}, they are parametrised by a fourth class,
\code{Graph}, which corresponds to the carrier of the algebraic
operations. In a standard, untyped setting, we would expect this
carrier to be just a set (a \code{Type}); the situation is slightly
more complicated here, since we define typed algebraic
structures. According to the previous explanations and
Fig.~\ref{fig:typed:algebras}, the \code{Graph} class encapsulates
several ingredients: a type for the set of indices (\code{T}), an
indexed family of types for the sets of morphisms (\code{X}), and for
each homset, an equivalence relation, \code{equal}---we cannot use
Leibniz equality: most models of Kleene algebra require a weaker
notion of equality (\code{relation} and \code{Equivalence} are
definitions from the standard library).

\begin{figure}[t]
  \begin{twolistings}
\begin{coq}
Class Graph := {
 T: Type;
 X: T -> T -> Type;
 equal: \forall n m, relation (X n m);
 equal_:> \forall n m, Equivalence (equal n m) }.

Class Monoid_Ops (G: Graph) := {
 dot: \forall n m p, X n m -> X m p -> X n p;
 one: \forall n,     X n n }.

Notation "x == y"  := (equal _ _ x y).
Notation "x @~y"    := (dot _ _ _ x y).
Notation "1"        := (one _). 
\end{coq}
&
\begin{coq}
Class SemiLattice_Ops (G: Graph) := {
 plus: \forall n m, X n m -> X n m -> X n m;
 zero: \forall n m, X n m }.
 
Class Star_Op (G: Graph) := {
 star: \forall n, X n n -> X n n }.
~
~

Notation "x + y"   := (plus _ _ x y).
Notation "0"       := (zero _ _). 
Notation "x#"      := (star _ x).
Notation "x <== y" := (x + y == y).
\end{coq}
  \end{twolistings}
  \caption{Classes for the typed algebraic operations.}
  \label{fig:classes:ops}
\end{figure}

We associate an intuitive notation to each operation, by using the
name provided by the corresponding class projection. To make the
effect of these definitions completely clear, assume that we have a
graph equipped with monoid operations (i.e., a typing context with %
\coqe{G: Graph} and \coqe{Mo: Monoid_Ops G}) and consider the
following proposition:
\begin{coq}
\forall (n m: T) (x: X n m) (y: X m n), x@y == 1.
\end{coq}
If we unfold notations, we get:
\begin{coq}
\forall (n m: T) (x: X n m) (y: X m n), equal _ _ (dot _ _ _ x y) (one _).
\end{coq}
Necessarily, by unification, the six placeholders have to be
filled as follows:
\begin{coq}
\forall (n m: T) (x: X n m) (y: X m n), equal n n (dot n m n x y) (one n).
\end{coq}
Now comes typeclass resolution: as explained in
Sect.~\ref{ssec:intro:tc}, the functions \code{T}, \code{X},
\code{equal}, \code{dot}, and \code{one}, which are class projections,
have implicit arguments that are automatically filled by typeclass
resolution (the graph instance for all of them, and the monoid
operations instance for \code{dot} and \code{one}). All in all, the
above concise proposition actually expands into:
\begin{coq}
\forall (n m: @@T G) (x: @@X G n m) (y: @@X G m n), @@equal G n n (@@dot G Mo n m n x y) (@@one G Mo n).
\end{coq}

\subsubsection*{Classes for algebraic laws.}

This was for syntax; we can finally define the classes for the laws
corresponding to the four algebraic structures we are interested
in. They are given in Fig.~\ref{fig:classes:laws}; we use the section
mechanism to assume a graph together with the operations, which become
parameters when we close the section. (We motivate our choice to have
separate classes for operations and for laws in
Sect.~\ref{sssec:tc:details:sep}.)

\begin{figure}[t]
\begin{coq}
Section.
 Context (G: Graph) {Mo: Monoid_Ops G} {SLo: SemiLattice_Ops G} {Ko: Star_Op G}.
 
 Class Monoid := {
   dot_compat:> \forall n m p, Proper (equal n m ==> equal m p ==> equal n p) (dot n m p);
   dot_assoc: \forall n m p q (x: X n m) (y: X m p) (z: X p q), x@(y@z) == (x@y)@z;
   dot_neutral_left:  \forall n m (x: X n m), 1@x == x;
   dot_neutral_right:  \forall n m (x: X m n), x@1 == x }.
 
 Class SemiLattice := {
   plus_compat:> \forall n m, Proper (equal n m ==> equal n m ==> equal n m) (plus n m);
   plus_neutral_left: \forall n m (x: X n m), 0+x == x;
   plus_idem: \forall n m (x: X n m), x+x == x;
   plus_assoc: \forall n m (x y z: X n m), x+(y+z) == (x+y)+z;
   plus_com: \forall n m (x y: X n m), x+y == y+x }.
  
 Class IdemSemiRing := {
   Monoid_:> Monoid;
   SemiLattice_:> SemiLattice;
   dot_ann_left:  \forall n m p (x: X m p), 0@x == (0: X m n);
   dot_ann_right: \forall n m p (x: X p m), x@0 == (0: X n m);
   dot_distr_left:  \forall n m p (x y: X n m) (z: X m p), (x+y)@z == x@z + y@z;
   dot_distr_right: \forall n m p (x y: X m n) (z: X p m), z@(x+y) == z@x + z@y }.
 
 Class KleeneAlgebra := {
   IdemSemiRing_:> IdemSemiRing;
   star_make_left: \forall n (x: X n n), 1+x#@x == x#;
   star_destruct_left: \forall n m (x: X n n) (y: X n m), x@y <== y -> x#@y <== y;
   star_destruct_right: \forall n m (x: X n n) (y: X m n), y@x <== y -> y@x# <== y }.
End.
\end{coq}
  \caption{Classes for the typed algebraic structures.}
  \label{fig:classes:laws}
\end{figure}

The \code{Monoid} class actually corresponds to the definition of a
category: we assume that composition (\code{dot}) is associative and
has \code{one} as neutral element. Its first field, \coqe{dot_compat},
requires that composition also preserves the user-defined equality: it
has to map equals to equals. (This field is declared with a special
symbol~(\coqe{:>}) and uses the standard \code{Proper} class, which is
exploited by Coq to perform rewriting with user-defined relations;
doing so adds \coqe{dot_compat} as a hint for typeclass resolution, so
that we can automatically rewrite in \code{dot} operands whenever it
makes sense.) Also note that since this class does not mention
semi-lattice operations nor the star operation, it does not depend on
\code{SLo} and \code{Ko} when we close the section.  We do not comment
on the \code{SemiLattice} class, which is quite similar.

The first two fields of \code{IdemSemiRing} implement the expected
inheritance relationship: an idempotent semiring is composed of a
monoid and a semi-lattice whose operations properly distribute. By
declaring these two fields with a \coqe{:>}, the corresponding
projections are added as hints to typeclass resolution, so that one
can automatically use any theorem about monoids or semi-lattices in
the context of a semiring. Note that we have to use type annotations
for the two annihilation laws: in both cases, the argument \code{n} of
$0$ (\code{zero}) cannot be inferred from the context, it has to be
specified.

Finally, we obtain the class for Kleene algebras by inheriting from
\code{IdemSemiRing} and requiring the three laws about Kleene star to
hold. The counterpart of \coqe{star_make_left} and the fact that
Kleene star is a proper morphism for \code{equal} are consequences of
the other axioms; this is why we do not include a \coqe{star_compat}
or \coqe{star_make_right} field in the signature: we prove these
lemmas separately (and we declare the former as an instance for
typeclass resolution), this saves us from additional proofs when
defining new models.

\medskip %
The following example illustrates the ease of use of this approach.
Here is how we would state and prove a lemma about idempotent
semirings:
\begin{coq}
Goal forall `{IdemSemiRing} n (x y: X n n), x@(y+1)+x == x@y+x.
Proof. 
  intros. 
  rewrite dot_distr_right, dot_neutral_right. (** (x@y+x)+x == x@y+x **)
  rewrite <-plus_assoc, plus_idem.
  reflexivity.
Qed.
\end{coq}
The special \coqe|`{IdemSemiRing}| notation allows us to assume a
generic idempotent semiring, with all its parameters (a graph, monoid
operations, and semi-lattice operations); when we use lemmas like
\coqe{dot_distr_right} or \coqe{plus_assoc}, typeclass resolution
automatically finds appropriate instances to fill their implicit
arguments. Of course, since such simple and boring goals occur
frequently in larger and more interesting proofs, we actually defined
high-level tactics to solve them automatically. For example, we have a
reflexive tactic called \coqe{semiring_reflexivity} which would solve
this goal directly: this is the counterpart to
\code{ring}~\cite{DBLP:conf/tphol/GregoireM05} for the equational
theory of typed, idempotent, non-commutative semirings.

\subsubsection*{Declaring new models.}
It remains to populate the above classes with concrete structures,
i.e., to declare models of Kleene algebra. We sketched the case of
heterogeneous binary relations and languages in
Fig.~\ref{fig:rel:lang:instances}; a user needing its own model of
Kleene algebra just has to declare it in the very same way. As
expected, it suffices to define a graph equipped with the various
operations, and to prove that they validate all the axioms. The
situation is slightly peculiar for languages, which form an
\emph{untyped} model: although the instances are parametrised by a set
\code{A} coding for the alphabet, there is no notion of
domain/co-domain of a language. In fact, all operations are total,
they actually lie in a one-object category where domain and co-domain
are trivial. Accordingly, we use the singleton type \code{unit} for
the index type \code{T} in the graph instance, and all operations just
ignore the superfluous parameters.
 
\begin{figure}[t]
  \begin{twolistings}
\begin{coq}
Definition rel A B := A -> B -> Prop.
Instance rel_G: Graph := {
  T := Type;
  X := rel;
  equal A B R S := \forall i j, R i j <-> S i j }.
Proof...
Instance rel_Mo: Monoid_Ops rel_G := {
  dot A B C R S := 
    fun (i: A)(j: C) => \exists k: B, R i k /\ S k j;
  one A := 
    fun (i j: A) => i=j }.
...
Instance rel_KA: KleeneAlgebra rel_G.
Proof...
\end{coq}
&
\begin{coq}
Definition lang A := list A -> Prop
Instance lang_G A: Graph := {
  T := unit;
  X _ _ := lang A;
  equal _ _ L K := \forall w, L w <-> K w }.
Proof...
Instance lang_Mo A: Monoid_Ops (lang_G A) := {
  dot _ _ _ L K := 
    fun w => \exists u v, w=u\appv /\ L u /\ K v;
  one _ := 
    fun w => w=[] }.
...
Instance lang_KA: KleeneAlgebra lang_G.
Proof...
\end{coq}
  \end{twolistings}
  \caption{Instances for heterogeneous binary relations and languages.}
  \label{fig:rel:lang:instances}
\end{figure}

\subsection{Reification: handling typed models.}
\label{ssec:untype:quote}

We also need to define a syntactic model in which to perform
computations: since we define a reflexive tactic, the first step is to
reify the goal (an equality between two expressions in an arbitrary
model) to use a syntactical representation.

For instance, suppose that we have a goal of the form %
\coqe{S@(R@S)#+f R == f R+(S@R)#@S}, where \code{R} and \code{S}
are binary relations and \code{f} is an arbitrary function on
relations. The usual methodology in Coq consists in defining a syntax
and an evaluation function such that this goal can be converted into
the following one:
\begin{center}
  \coqe{eval (var 1\odot(var 2\odot var 1)\ostr \oplus var 3) == eval (var 3 \oplus (var 1\odot var 2)\ostr\odot var 1)},
\end{center}
where \coqe{\oplus}, \coqe{\odot}, and \coqe{\ostr} are syntactic
constructors, and where \code{eval} implicitly uses a reification
environment, which corresponds to the following assignment:
\begin{center}
  \coqe|{1 \mapsto S; 2 \mapsto R; 3 \mapsto f R}|.
\end{center}

\subsubsection*{Typed syntax.}
The situation is slightly more involved here since we work with typed
models: \code{R} might be a relation from a set \code{A} to another
set \code{B}, \code{S} and \coqe{f R} being relations from \code{B} to
\code{A}. As a consequence, we have to keep track of domain/co-domain
information when we define the syntax and the reification
environments. The corresponding definitions are given in
Fig.~\ref{fig:typed:syntax}.
\begin{figure}[t]
\begin{twolistings}
\begin{coq}
Context `{KA: KleeneAlgebra}.
Variables src, tgt: label -> T.
Inductive reified: T -> T -> Type :=
| r_dot: \forall n m p, 
    reified n m -> reified m p -> reified n p
| r_one: \forall n, reified n n
| ...
| r_var: \forall i, reified (src i) (tgt i).
\end{coq}&
\begin{coq}
Variable env: forall i, X (src i) (tgt i).
Fixpoint eval n m (x: reified n m): X n m :=
  match x with
  | r_dot _ _ _ x y => eval x @~eval y
  | r_one _ => 1
  | ...
  | r_var i => env i
  end.
\end{coq} 
\end{twolistings}
\caption{Typed syntax for reification and evaluation function.}
\label{fig:typed:syntax}
\end{figure}
We assume an arbitrary Kleene algebra (in the previous example, it
would be the algebra of heterogeneous binary relations) and two
functions \code{src} and \code{tgt} associating a domain and a
co-domain to each variable (\code{label} is an alias for
\code{positive}, the type of positive numbers, which we use to index
variables). The \code{reified} inductive type corresponds to the typed
reification syntax: it has dependently typed constructors for all
operations of Kleene algebras, and an additional constructor for
variables, which is typed according to functions \code{src} and
\code{tgt}. To define the evaluation function, we furthermore assume
an assignation \code{env} from variables to elements of the Kleene
algebra with domain and co-domain as specified by \code{src} and
\code{tgt}.
Reifying a goal using this typed syntax is relatively easy: thanks to
the typeclass framework, it suffices to parse the goal, looking for
typeclass projections to detect operations of interest (recall for
example that a starred sub-term is always of the form %
\coqe{@@star _ _ _ _}, regardless of the current model---this model is
given in the first two placeholders). At first, we implemented this
step as a simple \code{Ltac} tactic. For efficiency reasons, we
finally moved to an OCaml implementation in a small plugin: this
allows one to use efficient data structures like hash-tables to
compute the reification environment, and to avoid type-checking the
reified terms at each step of their construction.

\subsubsection*{Untyped regular expressions.}

To build a reflexive tactic using the above syntax, we need a theorem
of the following form (keeping the reification environment implicit
for the sake of readability):
\begin{coq}
Theorem f_correct: forall n m (x y: reified n m), f x y = true -> eval x == eval y.
\end{coq}
The function \code{f} is the decision procedure; it works on reified
terms so that its type has to be %
\coqe{forall n m, reified n m -> reified n m -> bool}.
However, defining such a function directly would be rather
impractical: the standard algorithms underlying the decision procedure
are essentially untyped, and since these algorithms are rather
involved, extending them to take typed regular expressions into
account would require a lot of work.

Instead, we work with standard, untyped, regular expressions, as
defined by the inductive type \code{regex} from
Fig.~\ref{fig:regex}. Equality of regular expressions is defined
inductively, using the rules from equational logic and the laws of
Kleene algebra. By declaring the corresponding instances, we get an
untyped model (on the right-hand side of
Fig.~\ref{fig:erase:untype}---like for languages, we just ignore
domain/co-domain information).
\begin{figure}[t]
\begin{twolistings}
\begin{coq}
Inductive regex: Set := 
| dot:  regex -> regex -> regex 
| plus: regex -> regex -> regex 
| star: regex -> regex 
| one:  regex 
| zero: regex 
| var:  label -> regex.

Inductive eq: regex -> regex -> Prop :=
| eq_trans: forall y x z, x==y -> y==z -> x==z
| plus_idem: forall x, eq (x+x) x
| plus_compat: Proper (eq ==> eq ==> eq) plus
| star_make_left: forall x, eq (1+x#@x) (x#)
| ...
\end{coq}
&
\begin{coq}
Instance re_G: Graph := {
 T := unit;
 X _ _ := regex;
 equal _ _ := eq }.
Proof...

Instance re_Mo: Monoid_Ops re_G := {
 dot _ _ _ := dot;
 one _ := one }.

...   

Instance re_KA: KleeneAlgebra re_G.
Proof...
\end{coq}
\end{twolistings}
\caption{Regular expressions, axiomatic equality, and corresponding instances.}
\label{fig:regex}
\end{figure}
This is the main model we shall work with to implement the decision
procedure and prove its correctness (Sect.~\ref{sec:algo}): as
announced in Sect.~\ref{ssec:strat}, we will get:
\begin{coq}
Definition decide_kleene: regex -> regex -> bool := ...
Theorem Kozen94: forall x y: regex, decide_kleene x y = true <-> x == y.
\end{coq}
(Here the symbol \code{==} expands to the inductive equality
predicate \code{eq} from Fig.~\ref{fig:regex}.)

\subsubsection*{Untyping theorem.}

We still have to bridge the gap between this untyped decision
procedure (to be presented in Sect.~\ref{sec:algo}) and the
reification process we described for typed models.
To this end, we exploit a nice property of the equational theory of
typed Kleene algebra: it reduces to the equational theory of untyped
Kleene algebra~\cite{pous:csl10:utas}. In other words, a typed law
holds in all typed Kleene algebras whenever the underlying
untyped law holds in all Kleene algebras.

To state this result formally, it suffices to define the type-erasing
function \code{erase} from Fig.~\ref{fig:erase:untype}: this function
recursively removes all type decorations of a typed regular expression
to get a plain regular expression. The corresponding ``untyping
theorem'' is given on the right-hand side: two typed expressions whose
images under \code{erase} are equal in the model of untyped regular
expressions evaluate to equal values in any typed model, under any
variable assignation (again, the reification environment is left
implicit here).
\begin{figure}[t]
\begin{twolistings}
\begin{coq}
Fixpoint erase n m (x: reified n m): regex := 
  match x with 
  | r_dot _ _ _ x y => erase x @~erase y
  | r_one _ => 1
  | ...
  | r_var i => var i
  end.
\end{coq}&
\begin{coq}
Theorem erase_faithful: 
  forall n m (x y: reified n m), 
    erase x == erase y -> eval x == eval y.
Proof...


~
\end{coq}
\end{twolistings}
\caption{Type erasing function and untyping theorem.}
\label{fig:erase:untype}
\end{figure}
By composing this theorem with the correctness of the untyped decision
procedure---the previous theorem \code{Kozen94}, we get the following
corollary, which allows us to get a reflexive tactic for typed models
even though the decision procedure is untyped.
\begin{coq}
Corollary dk_erase_correct: forall n m (x y: reified n m), 
  decide_kleene (erase x) (erase y) = true -> eval x == eval y.
\end{coq}

Proving the untyping theorem is non-trivial, it requires the
definition of a proof factorisation system; see~\cite{pous:csl10:utas}
for a detailed proof and a theoretical study of other untyping
theorems. Also note that Kozen investigated a similar
problem~\cite{Koz98b} and came up with a slightly different solution:
he solves the case of the Horn theory rather than the equational
theory, at the cost of working in a restrained form of Kleene
algebras. He moreover relies on model-theoretic arguments, while our
considerations are purely proof-theoretic.

Finally note that as it is stated here, theorem \coqe{erase_faithful}
requires the axiom \coqe{Eqdep.eq_rect_eq} from Coq standard
library. This comes from the inductive type \code{reified} from
Fig.~\ref{fig:typed:syntax}, which has dependent parameters in an
arbitrary type (more precisely, the field \code{T} of an arbitrary
graph \code{G}). We get rid of this axiom in the library at the price
of an indirection: we actually make this inductive type depend on
positive numbers and we use an additional map to enumerate the
elements of \code{T} that are actually used (since terms are finite,
there are only finitely many such elements in a given goal). Since the
type of positive numbers has decidable equality, we can eventually
avoid using axiom \coqe{Eqdep.eq_rect_eq}~\cite{Hedberg98}.

\subsection{More details on our approach.}
\label{ssec:tc:details}

We conclude this section with additional remarks on the advantages and
drawbacks of our design choices; the reader may safely skip these and
move directly to Sect.~\ref{sec:mx}.

\subsubsection{Taking advantage of symmetry arguments.}
\label{sssec:tc:details:sym}

It is common practice in mathematics to rely on symmetry arguments to
avoid repeating the same proofs again and again.  Surprisingly, by
carefully designing our classes and defining appropriate instances, we
can also take advantage of some symmetries present in Kleene
algebra, in a formal and simple way.

The starting point is the following observation. Consider a typed
Kleene algebra as a category with additional structure on the homsets;
by formally reversing all arrows, we get a new typed Kleene
algebra. Therefore, any statement that holds in all typed Kleene
algebra can be reversed, yielding another universally true statement.
(This \emph{duality principle} is standard in category
theory~\cite{Maclane:cwm}; it is also used in lattice
theory~\cite{DaveyPriestley90}, where we can always consider the dual
lattice.)

\begin{figure}[t]
  \begin{twolistings}
\begin{coq}
Context {G: Graph} {Mo: Monoid_Ops G} 
         {SLo: SemiLattice_Ops G} 
         {Ko:  Star_Op G}.

Instance G': Graph := {
 T := T;
 X n m := X m n;
 equal n m := equal m n;
 equal_ n m := equal_ m n }.

Instance Mo': Monoid_Ops G' := {
 dot n m p x y := @@dot G Mo p m n y x;
 one := @@one G Mo }.

Instance SLo': SemiLattice_Ops G' := {
 plus n m := @@plus G SLo m n;
 zero n m := @@zero G SLo m n }.

Instance Ko': Star_Op G' := {
 star := @@star G Ko }.
\end{coq}
&
\begin{coq}
Instance M' {M: Monoid G}: Monoid G' := {
 dot_neutral_left n m := 
   @@dot_neutral_right G Mo M m n;
 dot_neutral_right n m := 
   @@dot_neutral_left G Mo M m n;
 dot_compat n m p x x' Hx y y' Hy := 
   @@dot_compat G Mo M p m n y y' Hy x x' Hx }.
Proof. 
 intros. symmetry. simpl. apply dot_assoc. 
Qed.

...

Instance KA' {KA: KleeneAlgebra G}: 
                          KleeneAlgebra G' := {
 star_destruct_left n m := 
   @@star_destruct_right G Mo SLo Ko KA m n;
 star_destruct_right n m := 
   @@star_destruct_left G Mo SLo KA m n }.
Proof...
\end{coq}
  \end{twolistings}
  \caption{Instances for the dual Kleene algebra.}
  \label{fig:dual}
\end{figure}

In Coq, it suffices to define instances corresponding to this dual
construction. These instances are given in Fig.~\ref{fig:dual}. The
dual graph and operations are obtained by swapping domains with
co-domains; we get composition by furthermore reversing the order of
the arguments. Proving that these reversed operations satisfy the laws
of a Kleene algebra is relatively easy since almost all laws already
come with their dual counterpart (we actually wrote laws with some
care to ensure that the dual operation precisely maps such laws to
their counterpart). The two exceptions are associativity of
composition, which is in a sense self-dual up to symmetry of equality,
and \coqe{star_make_left} whose dual is a consequence of the other
axioms, so that it was not included in the signature of Kleene
algebras--- Fig.~\ref{fig:classes:laws}. (Note that these instances
are dangerous from the typeclass resolution point of view: they
introduce infinite paths in the proof search trees. Therefore, we do
not export them and we use them only on a case by case basis.)

With these instances defined, suppose that we have proved
\begin{coq}
Lemma iter_right `{KA: KleeneAlgebra}: \forall n m x y (z: X n m), z@x <== y@z ~->~ z@x# <== y#@z.
\end{coq}
By symmetry we immediately get
\begin{coq}
Lemma iter_left `{KA: KleeneAlgebra}: \forall n m x y (z: X m n), x@z <== z@y  ~->~  x#@z <== z@y#.
Proof iter_right (KA:=KA').
\end{coq}
Indeed, instantiating the Kleene algebra with its dual in lemma
\coqe{iter_right} amounts to swapping domains and co-domains in the
type of variables (only \code{z} is altered since \code{x} and
\code{y} have square types) and reversing the order of all
products. Doing so, we precisely get the statement of lemma
\coqe{iter_left}, up to conversion.

By combining the above two lemmas, we finally get the following one,
which we actually use in Sect.~\ref{ssec:determ}.
\begin{coq}
Lemma iter `{KA: KleeneAlgebra}: \forall n m x y (z: X n m), x@z == z@y ~->~ x#@z == z@y#.
Proof...
\end{coq}

\subsubsection{Concrete structures.}
\label{sssec:tc:details:concrete}

Our typeclass-based approach may become problematic when dealing with
concrete structures without using our notations in a systematic
way. This might be a drawback for potential end-users of the library.
Indeed, suppose one wants to use a concrete type rather than our
uninformative projection \code{X} to quantify over some relation
\code{R} between natural numbers:
\begin{coq}
Check forall R: rel nat nat, R == R.
\end{coq}
This term does not type-check since Coq is unable to unify %
\coqe{rel nat nat} (the declared type for \code R) with %
\coqe{@@X _ _ _} (the type which is expected on both sides of a
\code{==}).  A solution in this case consists in declaring the
instance \coqe{rel_G} from Fig.~\ref{fig:rel:lang:instances} as a
``canonical structure'': doing so precisely tells Coq to use
\coqe{rel_G} when facing such a unification problem. (By the way, this
also tells Coq to use \coqe{rel_G} for unification problems of the
form \mbox{\coqe{Type}$\,=_\beta\,$\coqe{@@T _}}, which is required by
the above example as well.)

Unfortunately, this trick does not play well with our peculiar
representation of untyped models, like languages or regular
expressions (Fig.~\ref{fig:rel:lang:instances}
and~\ref{fig:regex}). Indeed, the dummy occurrences of \code{unit}
parameters prevent Coq from using the instance \coqe{lang_G} as a
canonical structure. Our solution in this case consists in using an
appropriate notation to hide the corresponding occurrences of \code{X}
behind an informative name: 
\begin{coq}
Notation language := (@@X lang_G tt tt).
Notation regex := (@@X re_G tt tt).
Check forall L: language, L#@L# == L#.
\end{coq}
Also note that the ability to declare more general hints for
unification~\cite{CoenT09} would certainly help to solve this problem
in a nicer way.

\subsubsection{Separation between operations an laws.}
\label{sssec:tc:details:sep}

When defining the classes for the algebraic structures, it might seem
more natural to package operations together with their laws. For
example, we could merge the classes \coqe{Monoid_Ops} and
\coqe{Monoid} from Fig.~\ref{fig:classes:ops}
and~\ref{fig:classes:laws}. There are at least two reasons for keeping
separate classes.

First, by separating operational contents from proof contents, we
avoid the standard problems due to the lack of proof irrelevance in
Coq, and situations where typeclass resolution might be
ambiguous. Indeed, having two proofs asserting that some operations
form a semiring is generally harmless; however, if we pack operations
with the proof that they satisfy some laws, then two distinct proofs
sometimes mean two different operations, which becomes highly
problematic. This would typically forbid the technique we presented
above to factorise some proofs by duality.

Second, this makes it possible to define other structures sharing the
same operations (and hence, notations), but not necessarily the same
laws.  We exploit this possibility, for example, to define a class for
Kleene algebra \emph{with converse} using fewer laws: the good
properties of the converse operation provide more symmetries so that
some laws become redundant (we use this class to get shorter proofs
for the instances from Fig.~\ref{fig:rel:lang:instances}: the models
of binary relations and languages both have a converse
operation). 

This choice is not critical for the library in its current state,
because we basically stop at Kleene algebra. However, based on
preliminary experiments, having this separation is crucial when
considering richer structures like residuated Kleene
lattices~\cite{Jipsen04} or allegories~\cite{FreydScedrov90}.

\section{Matrices.}
\label{sec:mx}

In this section, we describe our implementation of matrices, building
on the previously described framework. Matrices are indeed required to
formalise Kozen's initiality proof~\cite{Koz94b}, as explained in
Sect.~\ref{ssec:deka}.

\subsection{Which matrices to construct?}

Assume a graph \code{G}. There are at least three ways of defining a
new graph for matrices:
\begin{enumerate}[(1)]
\item Fix an object $\mathtt{u}\in\mathtt{T}$ and use natural numbers
  $(\mathbb N)$ as objects: morphisms between $n$ and $m$ are $n\times
  m$ matrices whose elements belong to the square homset \coqe{X u u}.
\item Use pairs $(\mathtt{u},n)\in\mathtt{T}\times\mathbb N$ as
  objects: morphisms from $(\mathtt{u},n)$ to $(\mathtt{v},m)$ are
  $n\times m$ matrices with elements in \code{X u v}.
\item Use lists $[\mathtt{u}_1,\dots,\mathtt{u}_n]\in\mathtt{T}^\star$
  as objects: a morphism from $[\mathtt{u}_1,\dots,\mathtt{u}_n]$ to
  $[\mathtt{v}_1,\dots,\mathtt{v}_m]$ is an $n\times m$ matrix $M$
  such that $M_{i,j}$ belongs to \code{X u$_i$ v$_j$}.
\end{enumerate}
The third option is the most theoretically appealing one: this is the
most general construction. Although we can actually build a typed
Kleene algebra of matrices in this way, this requires dealing with a
lot of dependent types, which can be tricky. The second option is also
rather natural from the mathematical point of view and it does not
impose a strongly dependent typing discipline.

However, while formalising the second or the third option is
interesting \emph{per se}, to get new models of typed Kleene algebras,
the first construction actually suffices for Kozen's initiality
proof. Indeed, this proof only requires matrices over regular
expressions and languages. Since these two models are untyped (their
type \code{T} for objects is just \code{unit}), the three
possibilities coincide (we can take \code{tt} for the fixed object
\code{u} without loss of generality). In the end, we chose the first
option, because it is the simplest one.

\subsection{Coq representation for matrices.}

According to the previous discussion, we assume a graph %
\coqe{G: Graph} and an object \coqe{u: T}. We furthermore abbreviate
the type \coqe{X u u} as \code{X}: this is the type of the
elements---sometimes called scalars.

\subsubsection*{Dependently typed representation.}
A matrix can be seen as a partial map from pairs of integers to
\code{X}, so that the Coq type for matrices could be defined as
follows:
\begin{coq}
Definition MX (n m: nat) := forall i j, i<n -> j<m -> X.
Definition mx_equal n m (M N: MX n m) i j (Hi: i<n) (Hj: j<m) := M i j Hi Hj == N i j Hi Hj.
\end{coq}
This corresponds to the dependent types approach: a matrix is a map to
\code{X} from two integers and two proofs that these integers are
lower than the bounds of the matrix. Except for the concrete
representation, this is the approach followed
in~\cite{BertotGBP08,pack:math:struct,color}. With such a type, every
access to a matrix element is made by exhibiting two proofs, to ensure
that indices lie within the bounds. This is not problematic for simple
operations like the function \coqe{mx_plus} below: it suffices to pass
the proofs around; this however requires more boilerplate for other
functions, like block decomposition operations.
\begin{coq}
Context {SLo: SemiLattice_Ops G}.
Definition mx_plus n m (M N: MX n m) i j (Hi: i<n) (Hj: j<m) := M i j Hi Hj + N i j Hi Hj.
\end{coq}

\subsubsection*{Infinite functions.}

\begin{figure}[t]
\begin{twolistings}
\begin{coq}
Context {SLo: SemiLattice_Ops G}.
Fixpoint sum i k (f: nat -> X) := 
  match k with 
  | O => O 
  | S k => f i + sum (S i) k f 
  end.
\end{coq}&
\begin{coq}
Context {Mo: Monoid_Ops G}.
Definition mx_dot n m p (M: MX n m) (N: MX m p) := 
  fun i j => sum O m (fun k => M i k @  N k j).

Definition mx_one n: MX n n :=
  fun i j => if eq_nat_bool i j then 1 else 0.
\end{coq}
\end{twolistings}
\caption{Definition of matricial product and identity matrix.}
\label{fig:mxdot:mxone}
\end{figure}

We actually adopt another strategy: we move bounds checks to equality
proofs, by working with the following definitions:
\begin{coq}
Definition MX n m := nat -> nat -> X.
Definition mx_equal n m (M N: MX n m) := forall i j, i<n -> j<m -> M i j == N i j.
\end{coq}
Here, a matrix is an infinite function from pairs of integers to
\code{X}, only equality is restricted to the actual domain of the
matrix. With these definitions, we do not need to manipulate proofs
when defining matrix operations, so that subsequent definitions are
easier to write.  For instance, the functions for matrix
multiplication and block manipulations are given in
Fig.~\ref{fig:mxdot:mxone} and Fig.~\ref{fig:mxsub:mxblock}. For
multiplication, we use a very naive function to compute the
appropriate sum: there is no need to provide an explicit proof that
each call to the functional argument is performed within the bounds.

\begin{figure}[t]
\begin{twolistings}
\begin{coq}
Definition mx_sub p q x y n m
  (M: MX p q): MX n m :=
  fun i j => M (x+i) (y+j).

Variables x y n m: nat.
Definition mx_sub00 := mx_sub (x\+n) (y\+m) 0 0 x y.
Definition mx_sub01 := mx_sub (x\+n) (y\+m) 0 y x m.
Definition mx_sub10 := mx_sub (x\+n) (y\+m) x 0 n y.
Definition mx_sub11 := mx_sub (x\+n) (y\+m) x y n m.
\end{coq}&
\begin{coq}
Definition mx_blocks x y n m 
  (M: MX x y) (N: MX x m)
  (P: MX n y) (Q: MX n m): MX (x\+n) (y\+m)
  := fun i j => match S i-x, S j-y with
      | O,   O   => M i j
      | O,   S j => N i j
      | S i, O   => P i j
      | S i, S j => Q i j
     end. 
\end{coq}
\end{twolistings}
\caption{Definition of sub-matrix extraction and block matrix construction.}
\label{fig:mxsub:mxblock}
\end{figure}

Similarly, the \coqe{mx_sub} function, for extracting a sub-matrix,
has a very liberal type: it takes an arbitrary $p\times q$ matrix $M$,
it returns an arbitrary $n\times m$ matrix, and this matrix is
obtained by reading $M$ from an arbitrary position $(x,y)$. This
function is then instantiated with more sensible arguments to get the
four functions corresponding to the decomposition of an
$(x+n)\times(y+m)$ matrix into four blocks. The converse function, to
define a matrix by blocks, is named \coqe{mx_blocks}.
%

Bounds checks are required a posteriori only, when proving properties
about these matrix operations, e.g., that multiplication is
associative or that the four sub-matrix functions preserve matricial
equality. This is generally straightforward: these proofs are done
within the interactive proof mode, so that bound checks can be proved
with high-level tactics like \coqe{omega}.
(Note that a similar behaviour could also be achieved with a
dependently typed definition of matrices by using Coq's \coqe{Program}
feature. We prefer our approach for its simplicity: \coqe{Program}
tends to generate large terms which are not so easy to work with.)

The correctness proof of our algorithm heavily relies on matricial
reasoning (Sect.~\ref{sec:algo}), and in particular block matrix
decomposition (Sect.~\ref{ssec:mxstar}
and~\ref{ssec:constr}). Despite this fact, we have not found major
drawbacks to this approach yet. We actually believe that it would
scale smoothly to even more intensive usages of matrices like, e.g.,\
linear algebra~\cite{gonthier:itp11}.

\subsubsection*{Phantom types.}
Unfortunately, these non-dependent definitions allow one to type the
following code, where the three additional arguments of \code{dot} are
implicit:
\begin{coq}
Definition ill_dot n p (M: MX n 16) (N: MX 64 p): MX n p := dot M N.
\end{coq}
This definition is accepted thanks to the conversion rule: the
dependent type \code{MX n m} does not mention \code{n} nor \code{m} in
its body, so that these arguments can be discarded by the type system
(we actually have \coqe{MX n 16 = MX n 64}). While such an ill-formed
definition will be detected at proof-time; it is a bit sad to loose
the advantages of a strongly typed programming language here.
We solved this problem at the cost of some syntactic sugar, by
resorting to an inductive singleton definition, reifying bounds in
\emph{phantom types}:
\begin{coq}
Record MX (n m: nat) := box { get: nat -> nat -> X }.
Definition mx_plus n m (M N: MX n m) := box n m (fun i j => get M i j + get N i j).
\end{coq}
Coq no longer equates types \code{MX n 16} and \code{MX n 64} with
this definition, so that the above \coqe{ill_dot} function is
rejected, and we can trust inferred implicit arguments (e.g., the
\code m argument of \code{dot}).

\subsubsection*{Computation.}
Although we do not use matrices for computations in this work, we also
advocate this lightweight representation from the efficiency point of
view.
First, using non-dependent types is more efficient: not a single
boundary proof gets evaluated in matrix computations.
Second, using functions to represent matrices allows for fine-grain
optimisation: it gives a lazy evaluation strategy by default, which
can be efficient if the matrix resulting of a computation is seldom
used, but we can also enforce a call-by-value behaviour for some
expressions, to avoid repeating numerous calls to a given expensive
computation. Indeed, we can define a \emph{memoisation} operator that
computes all elements of a given matrix, stores the results in a map,
and returns the closure that looks up in the map rather than
recomputing the result. The map can be implemented using lists or
binary trees, for example. In any case, we can then prove this
memoisation operator to be an identity so that it can be inserted in
matrix computations in a transparent way, at judicious places.

\begin{coq}
Definition mx_force n m (M: MX n m): MX n m := 
  let l := mx_to_maps M in box n m (fun i j => mget i (mget j l)). 
Lemma mx_force_id : forall n m (M : MX n m), mx_force M == M.
\end{coq}

\subsection{Taking the star of a matrix.}
\label{ssec:mxstar}

As expected, we declare the previous operations on matrices (e.g.,
Fig.~\ref{fig:mxdot:mxone}) as new instances, so that we can directly
use notations, lemmas, and tactics with matrices.  The type of these
instances are given below:
\begin{coq}
Instance mx_G: Graph := { T := nat; X := MX; equal := mx_equal }.
\end{coq}
\begin{coq}
Instance mx_SLo: SemiLattice_Ops G -> SemiLattice_Ops mx_G. 
Instance mx_Mo:~ SemiLattice_Ops G -> Monoid_Ops G -> Monoid_Ops mx_G. 
Instance mx_Ko:~ SemiLattice_Ops G -> Monoid_Ops G -> Star_Op G -> Star_Op mx_G. 
\end{coq}
\begin{coq}
Instance mx_SL:~ `{SemiLattice G} -> SemiLattice mx_G. 
Instance mx_ISR: `{IdemSemiRing G} -> IdemSemiRing mx_G. 
Instance mx_KA:~ `{KleeneAlgebra G} -> KleeneAlgebra mx_G. 
\end{coq}
To obtain the fourth and last instances, we have to define a star
operation on matrices, and show that it satisfies the laws for Kleene
star. We conclude this section about matrices by a brief description
of this construction---see~\cite{Koz94b} for a detailed proof.

\medskip

The idea is to proceed by induction on the size of the matrix: the
problem is trivial if the matrix is empty or of size $1\times 1$;
otherwise, we decompose the matrix into four blocks and we recurse as
follows~\cite{AhoHU74}:
\begin{align}
  \label{eq:mxstar}
  \tag{$\dag$}
  \left[\begin{array}{@{}c|c@{}} %
      A & B\\\hline C & D %
    \end{array}\right]^\star %
  &= %
  \left[\begin{array}{@{}c|c@{}} %
      A' & A'\cdot B\cdot D'\\\hline %
      D'\cdot C\cdot A' & %
      D'+D'\cdot C\cdot A'\cdot B\cdot D' %
    \end{array}\right] %
  &\text{where~}& %
  \begin{cases}
    D' = D^\star,\\ 
    A' = (A+B\cdot D'\cdot C)^\star
  \end{cases}
\end{align}
This definition may look mysterious; the special case where $C$ is
zero might be more intuitive:
\begin{align}
  \label{eq:mxstar:tri}
  \tag{$\ddag$}
  \left[\begin{array}{c|c} %
      A & B\\\hline 0 & D %
    \end{array}\right]^\star %
  &= %
  \left[\begin{array}{c|c} %
      A^\star & A^\star\cdot B\cdot D^\star\\\hline %
      0 & D^\star %
    \end{array}\right]\enspace. %
\end{align}
As long as we take square matrices for $A$ and $D$, the way we
decompose the matrix does not matter (we actually have to prove
it). In practice, since we work with Coq natural numbers (\code{nat}),
we choose $A$ of size $1\times 1$: this allows recursion to go
smoothly (if we were interested in efficient matrix computations, it
would be better to half the matrix size).

\begin{figure}[t]
\begin{twolistings}
\begin{coq}
Definition mx_star' x n 
  (sx: MX x x -> MX x x)
  (sn: MX n n -> MX n n) 
  (M: MX (x\+n) (x\+n)): MX (x\+n) (x\+n) :=
 let A := mx_sub00 M in
 let B := mx_sub01 M in
 let C := mx_sub10 M in
 let D := mx_sub11 M in
 let D' := sn D in
 let A' := sx (A+B@D'@C) in
   mx_blocks 
        A'        (A'@B@D') 
    (D'@C@A') (D'+D'@C@A'@B@D').
\end{coq}&
\begin{coq}
Definition mx_star_11 (M: MX 1 1): MX 1 1 := 
  fun _ _ => (M O O)#.
 
Fixpoint mx_star n: MX n n -> MX n n :=
  match n with
  | O => fun M => M
  | S n => mx_star' mx_star_11 (mx_star n)
  end.

Theorem mx_star_block x n (M: MX (x\+n) (x\+n)):
  mx_star (x\+n) M == 
    mx_star' (mx_star x) (mx_star n) M.
Proof...
\end{coq}    
\end{twolistings}
\caption{Definition of the star operation on matrices.}
\label{fig:mxstar}
\end{figure}

The corresponding code is given in Fig.~\ref{fig:mxstar}. We first
define an auxiliary function, \coqe{mx_star'}, which follows the above
definition by blocks~\eqref{eq:mxstar}, assuming two functions to
perform the recursive calls (i.e., to compute $A'$ and $D'$).  The
function \coqe{mx_star_11} computes the star of a $1\times 1$ matrix
by using the star operation on the underlying element. Using these two
functions, we get the final \coqe{mx_star} function as a simple
fixpoint.
The proof that this operation satisfies the laws of Kleene algebras is
complicated~\cite{Koz94b}; note that by making explicit the general
block definition with the auxiliary function \coqe{mx_star'}, we can
easily state theorem \coqe{mx_star_block}: equation~\eqref{eq:mxstar}
holds for each possible decomposition of the matrix.

\section{The algorithm and its proof}
\label{sec:algo}

We now focus on the heart of our tactic: the decision procedure and
the corresponding correctness proof. The algorithm we chose to
implement to decide whether two regular expressions denote the same
language can be decomposed into five steps:
\begin{enumerate}[(1)]
\item normalise both expressions to turn them into ``strict star
  form'';
\item build non-deterministic finite automata with epsilon-transitions
  (\eNFA);
\item remove epsilon-transitions to get non-deterministic finite
  automata (NFA);
\item determinise the automata to obtain deterministic finite automata
  (DFA);
\item check that the two DFAs are equivalent.
\end{enumerate}
The fourth step can produce automata of exponential size. Therefore,
we have to carefully select our construction algorithm, so that it
produces rather small automata. More generally, we have to take a
particular care about efficiency; this drives our choices about both
data structures and algorithms.

\begin{figure}[t]
  \begin{twolistings}
\begin{coq}
Module MAUT.                        
 Record t := mk {
  size:     nat; 
  initial:  MX 1 size;
  delta:    MX size size;
  final:    MX size 1 }.
 Definition eval(A: t): regex := 
  mx_to_scal (initial A@delta A#@final A).
End MAUT.



Module NFA.
 Record t := mk {
  size:     state;                  
  labels:   label;
  delta:    label -> state -> stateset; 
  initial:  stateset;                  
  final:    stateset }.
 Definition to_MAUT(A: t): MAUT.t := ...
 Definition eval := MAUT.eval \o to_MAUT.
End NFA.
\end{coq}
&
\begin{coq}
Module eNFA.
 Record t := mk {
  size:     state;  
  labels:   label;
  epsilon:  state -> stateset;      
  delta:    label -> state -> stateset; 
  initial:  state;                  
  final:    state }.
 Definition to_MAUT(A: t): MAUT.t := ...
 Definition eval := MAUT.eval \o to_MAUT.
End eNFA.

Module DFA.
 Record t := mk {
  size:     state;                  
  labels:   label;
  delta:    label -> state -> state; 
  initial:  state;                  
  final:    stateset }.
 Definition to_MAUT(A: t): MAUT.t := ...
 Definition eval := MAUT.eval \o to_MAUT.
End DFA.
\end{coq}
  \end{twolistings}
  \caption{Coq types and evaluation functions for the four automata representations.}
  \label{fig:types}
\end{figure}
The Coq types we used to represent finite automata are given in
Fig.~\ref{fig:types}; we use modules only for handling the name-space;
the type \regex{} is that from Fig.~\ref{fig:regex}
(Sect.~\ref{ssec:untype:quote}), \code{label} and \code{state} are
aliases for the type of numbers. The first record type, \code{MAUT.t},
corresponds to the matricial representation of automata; it is rather
high-level but computationally inefficient (\code{MX n m} is the type
of $n\times m$ matrices over \regex{}---Sect.~\ref{sec:mx}). We only
use this type in proofs, through the evaluation function
\code{MAUT.eval} (the function \coqe{mx_to_scal} casts a $1\times 1$
matrix into a regular expression). The three other types are efficient
representations for the three kinds of automata we mentioned above;
fields \code{size} and \code{labels} respectively code for the number
of states and labels, the other fields are self-explanatory. In each
case, we define a translation function to matricial automata,
\coqe{to_MAUT}, so that each kind of automata can eventually be
evaluated into a regular expression.

\begin{figure}[t]
  \centering
  \newcommand\A{} %
  \newcommand\ev{\textrm{eval}} %
  \begin{mathpar}
    \xymatrix @R=2.5em @C=2.5em {%
      \mathtt{regex}\ar[d]_{\text{~\hyperref[ssec:ssf]{1.} Normalisation}\quad}&
      x\ar@{|->}[d]\ar@/^1.5em/@{==}[d]^\A &&& y\ar@{|->}[d]\ar@/_1.5em/@{==}[d]_\A\\%
      \mathtt{regex}\ar[d]_{\text{\hyperref[ssec:constr]{2.} Construction}\quad}&
      x'\ar@{|->}[d]\ar@{==}[rd]^\A &&& y'\ar@{|->}[d]\ar@{==}[ld]_\A\\%
      \mathtt{eNFA.t}\ar[d]_{\text{\hyperref[ssec:epsilon]{3.} Epsilon removal}\quad}&
      A_1\ar@{|->}[d]\ar@{|->}[r]^\ev&\cdot\ar@{==}[d]^\A&\cdot\ar@{==}[d]_\A&B_1\ar@{|->}[l]_\ev\ar@{|->}[d]\\%
      \mathtt{NFA.t} \ar[d]_{\text{\hyperref[ssec:determ]{4.} Determinisation}\quad}&
      A_2\ar@{|->}[d]\ar@{|->}[r]^\ev&\cdot\ar@{==}[d]^\A&\cdot\ar@{==}[d]_\A&B_2\ar@{|->}[l]_\ev\ar@{|->}[d]\\%
      \mathtt{DFA.t}&A_3\ar@{|->}[r]^\ev&\cdot\ar@{==}[r]^\A&\cdot&B_3\ar@{|->}[l]_\ev
      \ar@{-}@/^1.7em/[lll]^{\text{\hyperref[ssec:equiv]{5.} Equivalence check}}%
    }%
 \end{mathpar}
 \caption{Overall picture for the algorithm and its correctness.}
 \label{fig:correction}
\end{figure}
The overall structure of the correctness proof is depicted in
Fig.~\ref{fig:correction}. Datatypes are recalled on the left-hand
side; the outer part of the right-hand side corresponds to
computations: starting from two regular expressions $x$ and
$y$, two DFAs $A_3$ and $B_3$ are constructed and tested for
equivalence. The proof corresponds to the inner equalities
(\code{==}): each automata construction preserves the semantics of the
initial regular expressions, two DFAs evaluate to equal values when
they are declared equivalent by the corresponding algorithm.

\medskip

In the following sections, we give more details about each step of the
decision procedure, together with a sketch of our correctness proof
(although we work with different algorithms, this proof is largely
based on Kozen's one~\cite{Koz94b}).

\subsection{Normalisation, strict star form}
\label{ssec:ssf}

There exists no complete rewriting system to decide equations of
Kleene algebra (their equational theory is not finitely
based~\cite{redko64}); this is why one usually goes through finite
automata constructions. One can still use rewriting techniques to
simplify the regular expressions before going into these expensive
constructions. By doing so, one can reduce the size of the generated
automata, and hence, the time needed to check for their equivalence.

For example, a possibility consists in normalising expressions with
respect to the following convergent rewriting system. (Although we
actually implemented this trivial optimisation, we will not discuss it
here.)
\begin{align*}
  x\cdot 0 &\to 0 & 0\cdot x&\to 0 & x+0 &\to x & 0+x\to x \\
  x\cdot 1 &\to x & 1\cdot x&\to x & 0^\star&\to 1
\end{align*}
Among other laws one might want to exploit in a preliminary
normalisation step, there are the following ones:
\begin{align*}
  1^\star&\to 1 &{x^\star}^\star &\to x^\star\enspace.
\end{align*}
More generally, any star expression $x^\star$ where $x$ accepts the
empty word can be simplified using the simple syntactic procedure
proposed by Br\"uggemann-Klein~\cite{klein93}. For example, this
procedure reduces the expression on the left-hand side below to the
one on the right-hand side, which is in \emph{strict star form}: all
occurrences of the star operation act on \emph{strict} regular
expressions, regular expressions that do not accept the empty word.
\begin{equation*}
  ((a+1)\cdot((b+1)^\star\cdot c +d^\star))^\star 
  \quad\to\quad (a+b^\star\cdot c+d)^\star\enspace.
\end{equation*}
In Coq, this procedure translates into a simple fixpoint whose
correctness relies on the following laws:
\begin{align*}
  (x+1)^\star &= x^\star\\
  (x+y^\star)^\star &= (x+y)^\star\\
  (x\cdot y)^\star &= (x+y)^\star \tag{if $x$ and $y$ accept the
    empty word}
\end{align*}
\begin{center}
\begin{coq}
Fixpoint ssf: regex -> regex := ...
Theorem ssf_correct: forall x, ssf x == x.
\end{coq}
\end{center}
The above theorem corresponds to the first step of the overall proof,
as depicted in Fig.~\ref{fig:correction}.
As we shall explain in Sect.~\ref{ssec:epsilon}, working with
expressions in strict star form also allows us to get a simpler and
more efficient algorithm to remove epsilon transitions. This means
that we also proved the \code{ssf} function complete, i.e., that it
always produces expressions in strict star form:

\begin{center}
\begin{coq}
Inductive strict_star_form: regex -> Prop := ...
Theorem ssf_complete: forall x, strict_star_form (ssf x).
\end{coq}
\end{center}

One could also normalise expressions modulo idempotence of $+$, to
avoid replications in the generated automata. This in turn requires
normalising terms modulo associativity and commutativity of $+$, and
associativity of $\cdot$, so that terms like $((a+b)\cdot c)\cdot
d+(b+a)\cdot (c\cdot d)$ can be reduced modulo idempotence. Such a
phase can easily be implemented, but it results in a slower procedure
in practice (normalisation requires quadratic time and non-trivial
instances of the idempotence law do not appear so frequently). We do
not include this step in the current release.

\subsection{Construction}
\label{ssec:constr}

There are several ways of constructing an \eNFA{} from a regular
expression.  At first, we implemented Thompson's
construction~\cite{thompson68}, for its simplicity; we finally
switched to a variant of Ilie and Yu's
construction~\cite{Ilie-yu-FollowAutomata}, which produces smaller
automata.  This algorithm constructs an automaton with a single
initial state and a single accepting state (respectively denoted by
$i$ and $f$); it proceeds by structural induction on the given regular
expression. The corresponding steps are depicted on the left-hand side
of Fig.~\ref{fig:construction}; the first drawing corresponds to the
base cases (zero, one, variable); the second one is union (plus): we
recursively build the two sub-automata between $i$ and $f$; the third
one is concatenation: we introduce a new state, $p$, build the first
sub-automaton between $i$ and $p$, and the second one between $p$ and
$f$; the last one is for iteration (star): we build the sub-automata
between a new state $p$ and $p$ itself, and we link $i$, $p$, and $f$
with two epsilon-transitions. The corresponding Coq code is given on
the right-hand side. To avoid costly union operations, we actually use
an accumulator (\code{A}) to which we recursively add states and
transitions (the functions \coqe{add_one} and \coqe{add_var}
respectively add epsilon and labelled transitions to the
accumulator---the function \code{incr} adds a new state to the
accumulator and returns this state together with the extended
accumulator).

\begin{figure}[t]
  \centering
  \begin{minipage}[t]{.35\linewidth}
    \begin{mathpar}
      \xymatrix @R=1em { %
        *+[o][F]{i}\ar[r]_{\emptyset/\epsilon/a} &
        *+[o][F]{f} }\\ %
      \xymatrix @R=1em { \\%
        *+[o][F]{i}\ar@/_2em/@{-}[r]\ar@/_1em/@{-}[r]_y %
                   \ar@/^2em/@{-}[r]\ar@/^1em/@{-}[r]^x & %
        *+[o][F]{f} %
      }\\ %
      \xymatrix @R=1em { %
        *+[o][F]{i}\ar@/_/@{-}[r]\ar@/^/@{-}[r]_x & %
        *+[o][F.]{p}\ar@/_/@{-}[r]\ar@/^/@{-}[r]_y & %
        *+[o][F]{f} %
      }\\ %
      \xymatrix @R=2em @C=2em { &*{}&\\
        *+[o][F]{i}\ar[r]_\epsilon & %
        *+[o][F.]{p}\ar[r]_\epsilon %
        \ar@{-}@(ul,ur)[]^x %
        \ar@{-}@(ul,l)[u] %
        \ar@{-}@(ur,r)[u] %
        & *+[o][F]{f} %
      } %
    \end{mathpar}
  \end{minipage}
  \qquad
  \begin{minipage}[t]{.55\linewidth}
\begin{coq}
Fixpoint build x i f A :=
  match x with 
    | zero     => A
    | one      => add_one i f A 
    | var a    => add_var a i f A
    | plus x y => build x i f (build y i f A)
    | dot x y  =>
        let (p,A) := incr A in
          build x i p (build y p f A)
    | star x   =>
        let (p,A) := incr A in 
          add_one i p (build x p p (add_one p f A))
  end.
\end{coq}    
  \end{minipage}
  \caption{Construction algorithm---a variant of Ilie and Yu's construction.}
\label{fig:construction}
\end{figure}

We actually implemented this algorithm twice, by using two distinct
datatypes for the accumulator: first, with a high-level matricial
representation; then with efficient maps for storing epsilon and
labelled transitions. Doing so allows us to separate the correctness
proof into an algebraic part, which we can do with the high-level
representation, and an implementation-dependent part consisting in
showing that the two versions are equivalent. 

These two versions correspond to the modules given in
Fig.~\ref{fig:construction:modules}.
Basically, we have the record types \code{MAUT.t} and \code{eNFA.t}
from Fig.~\ref{fig:types}, without the fields for initial and final
states. (The other difference being that we use maps rather than
functions on the the efficient side---\coqe{pre_eNFA}.) 
On the high-level side---\coqe{pre_MAUT}, we use generic matricial
constructions: adding a transition to the automaton consists in
performing an addition with the matrix containing only that transition
(\coqe{mx_point i f x} is the matrix with \code x at position
\code{(i,f)} and zeros everywhere else); adding a state to the
automaton consists in adding a empty row and a empty column to the
matrix, thanks to the \coqe{mx_blocks} function
(defined in Fig.~\ref{fig:mxsub:mxblock}).
We did not include the corresponding details for the low-level
representation: they are slightly verbose and they can easily be
deduced. Notice that \coqe{pre_NFA} does not include a generic
\code{add} function: while the matricial representation allows us to
label transitions with arbitrary regular expressions, the efficient
representation statically ensures that transitions are labelled either
with epsilon or with a variable (a letter of the alphabet).

The final construction functions, from \code{regex} to \code{MAUT.t}
or \code{eNFA.t}, are obtained by calling \code{build} between the two
states $0$ and $1$ of an empty accumulator (note that the occurrence
of $0$ in the definition of \coqe{pre_MAUT.empty} denotes the empty
$(2,2)$-matrix).

\begin{figure}
\begin{twolistings}
\begin{coq}
Module pre_MAUT.
 Record t := mk {
   size:  nat;
   delta: MX size size }.

 Definition to_MAUT i f A := MAUT.mk 
   (mx_point 0 i 1) (delta A) (mx_point f 0 1).
 Definition eval i f := MAUT.eval \o (to_MAUT i f)

 Definition add (x: regex) i f A :=
   mk _ (delta A + mx_point i f x)
 Definition add_one := add 1.
 Definition add_var a := add (var a).
 Definition incr A := let mk n M := A in 
   (n, mk (n+1) (mx_blocks M 0 0 0)).
 Fixpoint build x i f A := (* Fig. 15 *).

 Definition empty := mk 2 0.
 Definition regex_to_MAUT x :=
   to_MAUT 0 1 (build x 0 1 empty).
End pre_MAUT.
\end{coq}
&
\begin{coq}
Module pre_eNFA.
 Record t := mk {
   size:     state;
   labels:   label;
   epsmap:   statemap stateset;
   deltamap: statelabelmap stateset }.

 Definition to_eNFA i f A := ...



 Definition add_one i f A := ... 
 Definition add_var a i f A := ... 
 Definition incr A := ...

 Fixpoint build x := (* Fig. 15 *).

 Definition empty := mk 2 0 [] [].
 Definition regex_to_eNFA x :=
   to_eNFA 0 1 (build x 0 1 empty).
End pre_eNFA.
\end{coq}
\end{twolistings}
\caption{The two modules for the construction algorithm.}
\label{fig:construction:modules}
\end{figure}

Since the two versions of the algorithm only differ by their
underlying data structures, proving that they are equivalent is
routine (\coqe{[=]} denotes matricial automata equality):
\begin{coq}
Lemma constructions_equiv: forall x, regex_to_MAUT x [=] eNFA.to_MAUT (regex_to_eNFA x).
\end{coq}
Let us now focus on the algebraic part of the proof. We
have to show: 
\begin{coq}
Theorem construction_correct: forall x, MAUT.eval (regex_to_MAUT x) == x.
\end{coq}
The key lemma is the following one: calling \code{build x i f A} to
insert an automaton for the regular expression \code x between the
states \code i and \code f of \code A is equivalent to inserting
directly a transition with label \code x (recall that transitions can
be labelled with arbitrary regular expressions in matricial automata);
moreover, this holds whatever the initial and final states \code s and
\code t we choose for evaluating the automaton.
\begin{coq}
Lemma build_correct: forall x i f s t A, 
  i<size A -> f<size A -> s<size A -> t<size A ->
  eval s t (build x i f A) == eval s t (add x i f A).
\end{coq}
As expected, we proceed by structural induction on the regular
expression \code x. As an example of the involved algebraic reasoning,
the following property of star w.r.t.\ block matrices is used twice in
the proof of the above lemma: with $(x,y,z) = (e,0,f)$, it gives the
case of a concatenation $(e\cdot f)$; with $(x,y,z) = (1,e,1)$ it
yields iteration $(e^\star)$. This laws follows from the general
characterisation of the star operation on block matrices
(Equation~\eqref{eq:mxstar} in Sect.~\ref{ssec:mxstar}). In both
cases, the line and the column that are added on the left-hand side
correspond to the state $(p)$ generated by the construction.
\begin{small}
  \begin{align*}
    \left[\begin{array}{ccc|c}&u&&0\\\end{array}\right] 
    \cdot 
    \left[
      \begin{array}{ccc|r}
        &\vdots & & 0\\ \cdots & M_{i,f}& \cdots &
        x\\ & \vdots & &0 \\ \hline 0 & z & 0& y
      \end{array}
    \right]^\star 
    \cdot 
    \left[
      \begin{array}{c}
        \\[.6em]v\\[.6em]\\\hline0\\
      \end{array}
    \right] 
    &\quad=\quad
    u \cdot \left[
      \begin{array}{ccc} 
        &\vdots & \\ \cdots & M_{i,f} + x \cdot
        y^\star \cdot z & \cdots \\ & \vdots &
      \end{array}
    \right]^\star \cdot v
  \end{align*}
\end{small}  

\noindent
In the special case where \code A is the empty accumulator, lemma
\coqe{build_correct} gives:

\smallskip
\begin{center}
  \begin{tabular}{r@{\quad\code{==}\quad}l}
    \coqe|MAUT.eval (regex_to_MAUT x)|& %
    \coqe|eval 0 1 (build x 0 1 empty)|\\
    & \coqe|eval 0 1 (add x 0 1 empty)|\\[.2em]
    & $\left[\begin{array}{cc}1&0\\\end{array}\right]\cdot %
    \left[\begin{array}{cc}0&x\\0&0\\\end{array}\right]^\star\cdot %
    \left[\begin{array}{c}0\\1\\\end{array}\right]$\\[.9em]
    & $\left[\begin{array}{cc}1&0\\\end{array}\right]\cdot %
    \left[\begin{array}{cc}1&x\\0&1\\\end{array}\right]\cdot %
    \left[\begin{array}{c}0\\1\\\end{array}\right]$\\
    & \coqe|x| \\
  \end{tabular}
\end{center}
i.e., theorem \coqe{construction_correct}.

\noindent Finally, by combining the equivalence of the two algorithms
(lemma \coqe{constructions_equiv}) and the correctness of the
high-level one (theorem \coqe{construction_correct}), we obtain the
correctness of the efficient construction algorithm. In other words,
we can fill the two triangles corresponding to the second step in
Fig.~\ref{fig:correction}:
\begin{coq}
Theorem construction_correct': forall x, eNFA.eval (regex_to_eNFA x) == x.
\end{coq}

\subsection{Epsilon transitions removal}
\label{ssec:epsilon}

The automata obtained with the above construction contain
epsilon-transitions: each starred sub-expression produces two
epsilon-transitions, and each occurrence of $1$ gives one
epsilon-transition. Indeed, their transitions matrices are of the form
$M=J+N$ with $N=\sum_{a} a \cdot N_a$, where $J$ and the $N_a$ are 0-1
matrices. These matrices just correspond to the graphs of epsilon and
labelled transitions.

Removing epsilon-transitions can be done at the algebraic level using
the following law:
\begin{align*}
  (x+y)^\star &= x^\star\cdot\left(y \cdot x^\star\right)^\star\enspace,
\end{align*}
from which we get
\begin{align*}
  u\cdot(J+N)^\star\cdot v & = %
  u\cdot J^\star\cdot\left(N\cdot J^\star\right)^\star\cdot v\enspace,
\end{align*}
so that the automata $\tuple{u,M,v}$ and $\tuple{u\cdot J^\star,
  N\cdot J^\star,v}$ are equivalent. We can moreover notice that the
latter automaton no longer contains epsilon-transitions: this is a NFA
(the transition matrix, $N\cdot J^\star$, can be written as $\sum_{a}
a\cdot N_a\cdot J^\star$, where the $N_a\cdot J^\star$ are 0-1
matrices).

This algebraic proof is not surprising: looking at 0-1 matrices as
binary relations between states, $J^\star$ actually corresponds to the
reflexive-transitive closure of $J$.

\medskip

Although this is how we prove the correctness of this step, computing
$J^\star$ algebraically is inefficient: we have to implement a proper
transitive closure algorithm for the low-level representation of
automata. We actually rely on a property of the construction from
Sect.~\ref{ssec:constr}: when given regular expressions in strict star
form (Sect.~\ref{ssec:ssf}), the produced \eNFA{}s have acyclic
epsilon-transitions. Intuitively, the only possibility for introducing
an epsilon-cycle in the construction from Sect.~\ref{ssec:constr}
comes from star expressions. Therefore, by forbidding the empty word
to appear in such cases, we prevent the formation of epsilon-cycles.

Consider for example Fig.~\ref{fig:example:construction}, where we
have executed the construction algorithm of
Fig.~\ref{fig:construction} on two regular expressions (these are the
expressions from Sect.~\ref{ssec:ssf}---the right-hand side expression
is the strict star form of the left-hand side one).
\begin{figure}[t]
  \begin{tabular}[b]{ccc}
    $((a+1)\cdot((b+1)^\star\cdot c +d^\star))^\star$
    &\qquad\qquad
    & ~$(a+b^\star\cdot c+d)^\star$
    \\
    $\xymatrix @C=.8em @R=1em {
      &&&&\cdot\ar@{.>}@(ul,ur)^{b,\epsilon}
      \ar@{.>}@/^/[rrd]^\epsilon\\
      &&\cdot\ar@{.>}@/^/[rru]^\epsilon\ar@{.>}@/^/[rrd]^\epsilon
      &&&&\cdot\ar@/^1.3em/[lldd]^c\\
      &&&&\cdot\ar@(ur,dr)^d\ar@{.>}[d]^\epsilon\\
      \ar[r]&
      \cdot\ar@{.>}[rrr]_\epsilon&&&
      \cdot\ar@{.>}@/^1.3em/[lluu]^{a,\epsilon}
      \ar@{.>}[rrr]_\epsilon&&&\circledcirc}$
    &&
    $\xymatrix @C=.8em {
      &&\cdot\ar@(dl,ul)^b\ar@{.>}@/^/[rr]^\epsilon&&
      \cdot\ar@/^1.2em/[ld]^c\\
      \ar[r]&
      \cdot\ar@{.>}[rr]_\epsilon&&
      \cdot\ar@{.>}@/^1.2em/[lu]^\epsilon
      \ar@(dl,dr)_{a,d}\ar@{.>}[rr]_\epsilon&&\circledcirc}$
  \end{tabular}
  \caption{Running the construction algorithm on an expression and its
    strict star form.}
  \label{fig:example:construction}
\end{figure}
There are two epsilon-loops in the left hand-side automaton,
corresponding to the two occurrences of star that are applied to
non-strict expressions ($(b+1)^\star$ and the whole term). On the
contrary, in the automaton generated from the strict star form---the
second regular expression, the states belonging to these loops are
merged and the corresponding transitions are absent: the
epsilon-transitions form a directed acyclic graph (here, a tree).
 
This acyclicity property allows us to use a very simple algorithm to
compute the transitive closure. With respect to standard algorithms
for the general (cyclic) case, this algorithm is easier to implement
in Coq, slightly more efficient, and simpler to certify.
More concretely, we need to prove that the construction algorithm
returns \eNFA{}s whose reversed epsilon-transitions are well-founded,
when given expressions in strict star form:
\begin{coq}
Definition eNFA_well_founded A :=
  well_founded (fun i j => In i (eNFA.epsilon A j)).
Theorem construction_wf: forall x, 
  strict_star_form x -> eNFA_well_founded (regex_to_eNFA x).
\end{coq}
(Note that this proof is non-trivial.)
Our function to convert \eNFA{}s into NFAs takes such a well-founded
proof as an argument, and uses it to compute the reflexive-transitive
closure of epsilon-transitions:
\begin{coq}
Definition eNFA_to_NFA (A: eNFA.t): eNFA_well_founded A -> NFA.t := ...
\end{coq}
This step is easy to implement since we can proceed by well-founded
induction. In particular, there is no need to bound the recursion
level with the number of states, to keep track of the states whose
transitive closure is being computed to avoid infinite loops, or to
prove that a function defined in this way terminates. Note that we
still use memoisation, to take advantage of the sharing offered by the
directed acyclic graph structure. Also note that since this function
has to be executed efficiently, we use a standard Coq trick by Bruno
Barras to avoid the evaluation of the well-foundness proof: we guard
this proof with a large amount of constructors so that the actual
proof is never reached in practice.

We finally prove that the previous function returns an automaton whose
translation into a matricial automaton is exactly
$\tuple{u\cdot J^\star, N\cdot J^\star,v}$, so that the above
algebraic proof applies. This closes the third step in
Fig.~\ref{fig:correction}.
\begin{coq}
Theorem epsilon_correct: forall A (HA: eNFA_well_founded A), 
  NFA.eval (eNFA_to_NFA A HA) == eNFA.eval A.
\end{coq}

\subsubsection*{Comparison with Ilie and Yu's construction.}
Let us make a digression here to compare our construction algorithm
with the one proposed by Ilie and Yu~\cite[Algorithm~4,
p.~144]{Ilie-yu-FollowAutomata}.  The steps of the recursive
procedure, as presented in Fig.~\ref{fig:construction}, are exactly
the same; the only difference is that they refine the automaton by
merging some states and removing useless transitions:
\begin{enumerate}[(a)]
\item the state introduced in the \code{dot} case is removed when it
  is preceded or followed by a single epsilon-transition;
\item epsilon-cycles introduced in the \code{star} case are merged
  into a single state;
\item if at the end of the algorithm, the initial state only has one
  outgoing epsilon-transition, the initial state is shifted along this
  transition;
\item duplicated transitions are merged into a single one.
\end{enumerate}

\begin{wrapfigure}[6]{r}{0pt}%
  $\raisebox{3cm}[3cm]{
   \xymatrix @C=.8em @R=1.5em {& %
    \cdot\ar@(ul,ur)^b\ar@/^.7em/[d]^c\\
    \ar[r]& \cdot\ar@{.>}@/^.7em/[u]^\epsilon
    \ar@(dl,dr)_{a,d}\ar@{.>}[rr]_\epsilon&&\circledcirc}}$
\end{wrapfigure}
\noindent
For instance, running Ilie and Yu's construction on the right-hand
side expression of Fig.~\ref{fig:example:construction} yields the
automaton on the right. This automaton is actually smaller than the
one we generate: two states and two epsilon-transitions are removed
using~(a) and~(c). Moreover, thanks to optimisation~(b), Ilie and Yu
also get this automaton when starting from the left-hand side
expression, although this expression is not in strict star form.

We did not implement~(a) for two reasons: first, this optimisation is
not so simple to code efficiently (we need to be able to merge states
and to detect that only one epsilon transition reaches a given state),
second, it was technically involved to prove its correctness at the
algebraic level (recall that we need to motivate each step by some
matricial reasoning). Similarly, although step~(c) is easy to
implement, proving its correctness would require substantial
additional work.  On the contrary, our presentation of the algorithm
directly enforces~(d): the data structures we use systematically merge
duplicate transitions.

The remaining optimisation is~(b), which would be even harder to
implement and to prove correct than~(a). Fortunately, by working with
expressions in strict star form, the need for this optimisation
vanishes: epsilon-cycles cannot appear. In the end, although we
implement~(b) by putting expressions in strict star form first, the
only difference with Ilie and Yu's construction is that we do not
perform steps~(a) and~(c).

\subsection{Determinisation}
\label{ssec:determ}

%
Starting from a NFA $\tuple{u,M,v}$ with $n$ states, the
determinisation algorithm consists in a standard depth-first
enumeration of the subsets that are accessible from the set of initial
states. It returns a DFA $\tuple{\det u, \det M, \det v}$ with $\det
n$ states, together with a injective map $\rho$ from $[1..\det n]$ to
subsets of $[1..n]$. We sketch the algebraic part of the correctness
proof.  Let $X$ be the rectangular $(\det n,n)$ 0-1 matrix defined by
$X_{sj} \triangleq j \in \rho(s)$; the intuition is that $X$ is a
``decoding'' matrix: it sends states of the DFA to the characteristic
vectors of the corresponding subsets of the NFA. By a precise analysis
of the algorithm, we prove that the following commutation properties
hold:
\begin{align*}
  \det M \cdot X &= X\cdot M\quad(1) & \det u \cdot X &= u\quad(2) & \det v &= X \cdot v\quad(3)
\end{align*} 
Equation~$(1)$ can be read as follows: executing a transition in the
DFA and then decoding the result is equivalent to decoding the
starting state and executing parallel transitions in the
NFA. Similarly, $(2)$~states that the initial state of the DFA
corresponds to the set of initial states of the NFA, and $(3)$
assesses that the final states of the DFA are those containing at
least one accepting state of the NFA.

From~$(1)$, we deduce that $\mbox{$\det M$\,}^\star \cdot X = X\cdot
M^\star$ using the lemma \code{iter} from
Sect.~\ref{sssec:tc:details:sym}; we conclude with $(2,3)$:
\begin{align*}
  \det u \cdot \mbox{$\det M$\,}^\star \cdot \det v %
  = \det u \cdot \mbox{$\det M$\,}^\star \cdot X\cdot v %
  = \det u \cdot X \cdot M^\star \cdot v %
  = u \cdot M^\star \cdot v \enspace.
\end{align*}
The DFA evaluates like the starting NFA: we can fill the two squares
corresponding to the fourth step in Fig.~\ref{fig:correction}.

\medskip

Let us mention a Coq-specific technical difficulty in the concrete
implementation of this algorithm. The problem comes from termination:
even though it theoretically suffices to execute the main loop at most
$2^n$ times (there are $2^n$ subsets of $[1..n]$), we cannot use this
bound directly in practice. Indeed, NFAs with 500 states frequently
result in DFAs of about a thousand states, which we should be able to
compute easily. However, using the number $2^n$ to bound the recursion
depth in Coq requires to compute this number before entering the
recursive function. For $n=500$ this is obviously out of reach (this
number has to be in unary format---\code{nat}---since it is used to
ensure structural recursion).

We have tried to use well-founded recursion, which was rather
inconvenient: this requires mixing some non-trivial proofs with the
code. We currently use the following ``pseudo-fixpoint operators'',
defined in continuation passing style:
\begin{coq}
Variables A B: Type.
Fixpoint linearfix n (f: (A -> B) -> A -> B) (k: A -> B) (a: A): B :=
  match n with O => k a | S n => f (linearfix n f k) a end.
Fixpoint powerfix n (f: (A -> B) -> A -> B) (k: A -> B) (a: A): B :=
  match n with O => k a | S n => f (powerfix n f (powerfix n f k)) a end.
\end{coq}
Intuitively, \code{linearfix n f k} lazily approximates a potential
fixpoint of the functional \code{f}: if a fixpoint is not reached
after \code{n} iterations, it uses \code{k} to escape. The
\code{powerfix} operator behaves similarly, except that it escapes
after $2^n-1$ iterations: we prove that \code{powerfix n f k a} is
equal to \code{linearfix ($2^n-1$) f k a}.
Thanks to these operators, we can write the code to be executed using
\code{powerfix}, while keeping the ability to reason about the simpler
code obtained with a naive structural iteration over $2^n$: both
versions of the code are easily proved equivalent, using the
intermediate \code{linearfix} characterisation.

\subsection{Equivalence checking}
\label{ssec:equiv}

Two DFAs are equivalent if and only if their respective minimised DFAs
are equal up-to isomorphism. Therefore, computing the minimised DFAs
and exploring all state permutations is sufficient to obtain
decidability.

\begin{figure}[t]
  \begin{align*}
    \xymatrix @C=.8em @R=4em {
      \ar[r]&
      *++[o][F] {x}\ar[rr]^a\ar@{-}[d]_1&& 
      *++[o][F=]{y}\ar@/^/[rr]^a\ar@{-}[d]_2&& 
      *++[o][F] {z}\ar@/^/[ll]^a\ar@{-}[d]_3\\ 
      \ar[r]&
      *++[o][F] {u}\ar[rr]_a&&
      *++[o][F=]{v}\ar@/^/[rr]^a&& 
      *++[o][F] {w}\ar@/^/[ll]^a
    }     
    &&
    \xymatrix @C=.9em @R=.9em {
      \ar[r]&
      *++[o][F] {x}\ar[rr]^{a,b}\ar@{-}[dddd]_1&& 
      *++[o][F=]{y}\ar[rr]^{a,b}\ar@{-}[ddd]_2 \ar@{.}[rrddd] && 
      *++[o][F=]{z}\ar@(ur,dr)^{a,b}\ar@{-}[ddd]_3 \ar@{-}[llddd]_(.3)4\\\\\\
      &&&
      *++[o][F=]{v}\ar@/^/[rr]^{a,b}&& 
      *++[o][F=]{w}\ar@/^/[ll]^(.6){a,b}\\
      \ar[r]& *++[o][F] {u}\ar[rru]_a\ar@/_1.1em/[rrrru]_b&&
   }
  \end{align*}
  \caption{Checking for DFA equivalence (Hopcroft and Karp).}
  \label{fig:fly}
\end{figure}
However, there is a more direct and efficient approach that does not
require minimisation: one can use the almost linear algorithm by
Hopcroft and Karp~\cite{HopcroftKarp,AhoHU74}. This algorithm proceeds
as follow: starting from two DFAs $\tuple{u_1,M_1,v_1}$ and
$\tuple{u_2,M_2,v_2}$, it first computes the disjoint union automaton
$\tuple{u,M,v}$, defined by
\begin{align*}
  u&=\left[\begin{array}{cc}u_1&u_2\\\end{array}\right] &
  M&=\left[\begin{array}{cc}M_1&0\\0&M_2\\\end{array}\right] &
  v&=\left[\begin{array}{c}v_1\\v_2\\\end{array}\right]\enspace.
\end{align*}
It then checks that the former initial states are equivalent by
coinduction.  Intuitively, two states are equivalent if they can match
each other's transitions to reach equivalent states, with the
constraint that no accepting state can be equivalent to a
non-accepting one.

Let us execute this algorithm on the simple example given on the
left-hand side of Fig.~\ref{fig:fly}. We start with the pair of states
$(x,u)$; these two states are non-accepting so that we can declare
them equivalent \emph{a priori}. We then have to check that they can
match each other's transitions, i.e., that $y$ and $v$ are
equivalent. Both states are accepting, we declare them equivalent, and
we move to the pair $(z,w)$ (according to the transitions of the
automata). Again, since these two states are non-accepting, we declare
them equivalent and we follow their transitions. This brings us back
to the pair $(y,v)$. Since this pair was already encountered, we can
stop: the two automata are equivalent, they recognise the same
language. 
The algorithm always terminates: there are finitely many pairs of
states, and each pair is visited at most once.

This presentation of the algorithm makes it quadratic in worst
case. Almost linear time complexity is obtained by recording a set of
equivalence classes rather than the set of visited pairs. To
illustrate this idea, consider the example on the right-hand side of
Fig.~\ref{fig:fly}: starting from the pair $(x,u)$ and following
transitions along $a$, we reach a situation where the pairs $(x,u)$,
$(y,v)$, $(z,w)$, and $(z,v)$ have been declared as equivalent and
where we still need to check transitions along $b$. All of them result
in already declared pairs, except the initial one $(x,u)$, which
yields $(y,w)$. Although this pair was not visited, it belongs to the
equivalence relation generated by the previously visited
pairs. Therefore, there is no need to add this pair, and the algorithm
can stop immediately. This makes the algorithm almost linear: two
equivalence classes are merged at each step of the loop so that this
loop is executed at most $n+m$ times, where $n$ and $m$ are the number
of states of the compared DFAs. Using a disjoint-sets data structure
for maintaining equivalence classes ensures that each step is done in
almost-constant time~\cite{cormen_introduction_2001}.

To our knowledge, there is only one implementation of disjoint-sets in
Coq~\cite{Leroy-backend}. However, this implementation uses \code{sig}
types to ensure basic invariants along computations, so that reduction
of the corresponding terms inside Coq is not optimal: useless proof
terms are constantly built and thrown away. Although this drawback
disappears when the code is extracted (the goal
in~\cite{Leroy-backend} was to obtain a certified compiler, by
extraction), this is problematic in our case: since we build a
reflexive tactic, computations are performed inside Coq.
Conchon and Filli\^atre also certified a persistent union-find data
structure in Coq~\cite{persistant_union_find}, but this development
consists in a modelling of an OCaml library, not in a proper Coq
implementation that could be used to perform computations.
Therefore, we had to re-implement and prove this data structure from
scratch. Namely, we implemented disjoint-sets
forests~\cite{cormen_introduction_2001} with path compression and the
usual ``union by rank'' heuristic, along the lines
of~\cite{Leroy-backend}, but without using \code{sig}-types.

We do not give the Coq code for checking equivalence of DFAs here: it
closely follows~\cite{AhoHU74} and can be downloaded
from~\cite{kleenecoq:web}. Note that since recursion is not
structural, we need to explicitly bound the recursion depth. As
explained above, the size of the disjoint union automaton $(n+m)$ does
the job.

\medskip

Like previously, the correctness of this last step reduces to
algebraic reasoning. Define a 0-1 matrix $Y$ to encode the equivalence
relation on states obtained with a successful run of the algorithm:
\begin{align*}
  Y_{ij}&=
  \begin{cases}
    1&\text{if states $i$ and $j$ are equivalent,}\\
    0&\text{otherwise.}
  \end{cases}
\end{align*}
We prove that this matrix satisfies the following properties (like for
the determinisation step, these proofs are quite technical and
correspond to a detailed analysis of the algorithm---in particular, we
have to show that the bound we impose for the recursion depth is
appropriate):

\begin{mathpar}
  1 \leq Y \quad(1)\and
  Y \cdot Y \leq Y \quad(2)\and
  Y \cdot M \leq M \cdot Y \quad(3) \\
  [u_1~0] \cdot Y = [0~u_2] \cdot Y\quad(4) \and
  Y \cdot v = v\quad(5) 
\end{mathpar} 
Equations $(1,2)$ correspond to the fact that $Y$ encodes a reflexive
and transitive relation. 
Equation $(3)$ comes from the fact that $Y$ is a simulation:
transitions starting from related states yield related states. The
last two equations assess that the starting states are related $(4)$,
and that related states are either accepting or non-accepting~$(5)$.

This allows us to conclude using algebraic reasoning: from $(1,2,3)$
and Kleene algebra laws, we deduce
\begin{align*}
  M^\star \cdot Y &= Y \cdot (M\cdot Y)^\star\enspace.\tag{6}
\end{align*}
Also notice that as a special case of~\eqref{eq:mxstar:tri}, we have
\begin{align*}
  M^\star =
  \left[\begin{array}{cc}M_1&0\\0&M_2\\\end{array}\right]^\star = 
  \left[\begin{array}{cc}M_1^\star&0\\0&M_2^\star\\\end{array}\right]\enspace,
\end{align*}
so that we have
$u_1 \cdot M_1^\star \cdot v_1 = [u_1~0] \cdot M^\star \cdot v$ and %
$u_2 \cdot M_2^\star \cdot v_2 = [0~u_2] \cdot M^\star \cdot v$. 
Correctness follows:
\begin{align*}
    u_1 \cdot M_1^\star \cdot v_1 %
  =~& [u_1~0] \cdot M^\star \cdot v\\
  =~& [u_1~0] \cdot M^\star \cdot Y \cdot v \tag{by 5}\\
  =~& [u_1~0] \cdot Y \cdot (M\cdot Y)^\star \cdot v \tag{by 6}\\
  =~& [0~u_2] \cdot Y \cdot (M\cdot Y)^\star \cdot v \tag{by 4}\\
  =~& [0~u_2] \cdot M^\star \cdot Y \cdot v \tag{by 6}\\
  =~& [0~u_2] \cdot M^\star \cdot v \tag{by 5} %
  = u_2 \cdot M_2^\star \cdot v_2\enspace.
\end{align*}
In other words, we obtained the bottom line equality of
Fig.~\ref{fig:correction}.

\subsection{Putting it all together}
\label{ssec:alltogether}

By combining the proofs from the above sections according to
Fig.~\ref{fig:correction}, we obtain the decision procedure and its
correctness proof:
\begin{coq}
Definition regex_to_DFA x :=
  let x' := ssf x in
  let A1 := regex_to_eNFA x' in
  let A2 := eNFA_to_NFA A1 (construction_wf (ssf_complete x)) in
  let A3 := NFA_to_DFA A2 in
    A3.
Definition decide_kleene x y := DFA_equiv (regex_to_DFA x) (regex_to_DFA y).  
Theorem decide_kleene_correct: forall x y, decide_kleene x y = true -> x == y.
\end{coq}
As explained in Sect.~\ref{ssec:untype:quote}, although the above
equality lies in the syntactic model of regular expressions, we can
actually port it to any model of typed Kleene algebras using
reification and the untyping theorem.

\subsection{Completeness: counter-examples}
\label{ssec:completeness}

As announced in Sect.~\ref{ssec:strat}, we also proved the converse
implication, i.e., \emph{completeness}. This basically amounts to
exhibiting a counter-example in the case where the DFAs are not
equivalent. From the algorithmic point of view, it suffices to record
the word that is being read in the algorithm from
Sect.~\ref{ssec:equiv}; when two states that should be equivalent
differ by their accepting status, we know that the current word is
accepted by one DFA and not by the other one. Accordingly, the \dk{}
function actually returns an \code{option (list label)} rather than a
Boolean, so that the counter-example can be given to the user---in
particular, in the above statement of \coqe{decide_kleene_correct},
the constant \code{true} should be replaced by \code{None}.
We can then get the converse of \coqe{decide_kleene_correct}:
\begin{coq}
Theorem decide_kleene_complete: forall x y w, decide_kleene x y = Some w -> \not(x == y).
\end{coq}
The proof consists in showing that the word \code{w} possibly returned
by the equivalence check algorithm is actually a counter-example, and
that the language accepted by a DFA is exactly the language obtained
by interpreting the regular expression returned by \code{DFA.eval}:
\begin{coq}
Definition DFA_language: DFA.t -> language := ...
Definition regex_language: regex -> language := ...
Lemma language_DFA_eval: forall A, DFA_language A == regex_language (DFA.eval A).
\end{coq}
(Recall that languages---predicates over lists of letters---form a
Kleene algebra which we defined in Fig.~\ref{fig:rel:lang:instances};
in particular, the above symbol \code{==} denotes equality in this
model, i.e., pointwise equivalence of the predicates.)  The function
\code{DFA.eval} corresponds to a matricial product
(Fig.~\ref{fig:types}) so that the above lemma requires us to work
with matrices over languages. This is actually the only place in the
proof where we need this model.

\section{Efficiency}
\label{sec:perfs}

Thanks to the efficient reduction mechanism available in
Coq~\cite{GregoireL02}, and since we carefully avoided mixing proofs
with code, the tactic returns instantaneously on typical use cases.
We had to perform some additional tests to check that the decision
procedure actually scales on larger expressions. This would be
important, for example, in a scenario where equations to be solved by
the tactic are generated automatically, by an external tool.

A key factor is the concrete representation of numbers, which we
detail first.

\subsection{Numbers, finite sets, and finite maps}
\label{ssec:datastructures}

To code the decision procedure, we mainly needed natural numbers,
finite sets, and finite maps.
Coq provides several representations for natural numbers: Peano
integers (\coqe{nat}), binary positive numbers (\coqe{positive}), and
big natural numbers in base $2^{31}$ (\coqe{BigN.t}), the latter being
shipped with an underlying mechanism to use machine integers and
perform efficient computations. (On the contrary, unary and binary
numbers are allocated on the heap, as any other datatype.)
Similarly, there are various implementations of finite maps and finite
sets, based on ordered lists (\coqe{FMapList}), AVL trees
(\coqe{FMapAVL}), or uncompressed Patricia trees
(\coqe{FMapPositive}).

While Coq standard library features well-defined interfaces for finite
sets and finite maps, the different definitions of numbers lack this
standardisation. In particular, the provided tools vary greatly
depending on the implementation. For example, the tactic \coqe{omega},
which decides Presburger's arithmetic on \coqe{nat}, is not available
for \coqe{positive}.
To abstract from this choice of basic data structures, and to obtain a
modular code, we designed a small interface to package natural numbers
together with the various operations we need, including sets and
maps. We specified these operations with respect to \coqe{nat}, and we
defined several automation tactics. In particular, by automatically
translating goals to the \coqe{nat} representation, we can use the
\coqe{omega} tactic in a transparent way.

We defined several implementations of this interface, so that we could
experiment with the possible choices and compare their performances.
Of course, unary natural numbers behave badly since they bring an
additional exponential factor. However, thanks to the efficient
implementation of radix-2 search trees for finite maps and finite sets
(\code{FMapPositive} and \code{FSetPositive}), we actually get higher
performances by using \code{positive} binary numbers rather than
machine integers (\code{BigN.t}). This is no longer true with the
extracted code: using machine integers is faster on large expressions
with a thousand internal nodes. 

It would be interesting to rework our code to exploit the efficient
implementation of persistent arrays in experimental versions of
Coq~\cite{ArmandGST10}. We could reasonably hope to win an order of
magnitude by doing so; this however requires a non-trivial interfacing
work since our algorithms were written for dynamically extensible maps
over unbounded natural numbers while persistent arrays are of a fixed
size, and over cyclic 31 bits integers.

\subsection{Benchmarks}
\label{ssec:benchs}

Two alternative certified decision procedures for regular expression
equivalence have been developped since we proposed the present
one; both of them rely on a simple algorithm based on Brzozowski's
derivatives~\cite{Brzozowski64,Rutten98}:
\begin{itemize}
\item Krauss and Nipkow~\cite{Krauss} implemented a tactic for
  Isabelle/HOL;
\item Coquand and Siles~\cite{CoquandS11} implemented their algorithm
  in Coq;
  they use a particularly nice induction scheme for finite sets, which
  is one of their main contributions.
\end{itemize}
\noindent
We performed some benchmarks to compare the performances of these two
implementations with ours (we leave the comparison of our approaches
for the related works section, Sect.~\ref{ssec:related}). The timings
are given in Table~\ref{tbl:benchmarks}, they have been obtained as
follows.

For each pair $(n,v)$ given in the first two columns, we generated 500
pairs of regular expressions, with exactly $n$ nodes and at most $v$
distinct variables\footnote{these pairs are available on the web for
  the interested reader~\cite{kleenecoq:web}.}. Since two random
expressions tend to always be trivially distinct, we artificially
modified these pairs to make them equivalent, by adding the full
regular expression on both sides.  For instance, the pair
$(a+b^\star,~a\cdot b\cdot c)$, with four nodes and three variables,
is turned into the pair $(a+b^\star+(a+b+c)^\star,~a\cdot b\cdot
c+(a+b+c)^\star)$. By doing so, we make sure that all algorithms
actually explore the whole DFAs corresponding to the initial
expressions. 

For each of these modified pairs, we measured the time required by
each implementation (CoSi, KrNi, and BrPo respectively stand for
Coquand and Siles' implementation, Krauss and Nipkow' one, and ours).
The timings were measured on a Macbook pro (Intel Core 2 Duo, 2.5GHz,
4Go RAM) running Mac OS X 10.6.7, with Coq 8.3 and Isabelle
2011-1. All times are given in seconds, they correspond to the tactic
scenario, where execution takes place inside Coq or Isabelle. (When
extracting our Coq procedure to OCaml, the resulting code executes
approximately 20 times faster.)

\begin{table}
  \small
  \renewcommand\skip{\multicolumn{1}{c|}{--}}
  \begin{tabular}{|c|c|c|r|r|r|r|r|}
  \hline
  nodes & vars & algo. & mean & 50\% & 90\% & 99\% & 100\% \\
  \hline
  \hline
  \multirow{3}{*}{5}&
  \multirow{3}{*}{2}&
     CoSi &      0.017 &      0.014 &      0.028 &      0.070 &      0.190 \\
  && KrNi &      0.067 &      0.066 &      0.071 &      0.091 &      0.097 \\
  && BrPo &      0.001 &      0.001 &      0.002 &      0.002 &      0.011 \\
  \hline
  \multirow{3}{*}{10}&
  \multirow{3}{*}{2}&
     CoSi &      0.070 &      0.047 &      0.136 &      0.516 &      1.456 \\
  && KrNi &      0.072 &      0.070 &      0.076 &      0.104 &      0.131 \\
  && BrPo &      0.002 &      0.002 &      0.002 &      0.003 &      0.004 \\
  \hline
  \multirow{3}{*}{20}&
  \multirow{3}{*}{2}&
     CoSi &      2.037 &      0.326 &      2.925 &     44.101 &    145.875 \\
  && KrNi &      0.134 &      0.135 &      0.152 &      0.166 &      0.534 \\
  && BrPo &      0.003 &      0.003 &      0.004 &      0.006 &      0.007 \\
  \hline
  \multirow{3}{*}{20}&
  \multirow{3}{*}{4}&
     CoSi &  $>$31.043 &     13.708 &     42.983 &    261.022 & $>$3600.000 \\
  && KrNi &      0.132 &      0.122 &      0.160 &      0.171 &      0.685 \\
  && BrPo &      0.006 &      0.006 &      0.008 &      0.012 &      0.016 \\
  \hline
  \multirow{3}{*}{50}&
  \multirow{3}{*}{4}&
     CoSi &      \skip &      \skip &      \skip &      \skip &      \skip \\
  && KrNi &      0.251 &      0.236 &      0.294 &      1.061 &      2.337 \\
  && BrPo &      0.019 &      0.018 &      0.028 &      0.049 &      0.057 \\
  \hline
  \multirow{3}{*}{100}&
  \multirow{3}{*}{10}&
     CoSi &      \skip &      \skip &      \skip &      \skip &      \skip \\
  && KrNi &      0.686 &      0.487 &      0.976 &      6.401 &      8.468 \\
  && BrPo &      0.135 &      0.128 &      0.190 &      0.314 &      0.359 \\
  \hline
  \multirow{3}{*}{200}&
  \multirow{3}{*}{20}&
     CoSi &      \skip &      \skip &      \skip &      \skip &      \skip \\
  && KrNi &  $>$30.420 &      2.123 &     16.026 &    1621.541 & $>$3600.000 \\
  && BrPo &      0.695 &      0.662 &      0.948 &      1.320 &      1.672 \\
  \hline
  \multirow{3}{*}{500}&
  \multirow{3}{*}{50}&
     CoSi &      \skip &      \skip &      \skip &      \skip &      \skip \\
  && KrNi & $>$280.340 &     96.220 & $>$900.000 & $>$3600.000 & $>$4703.633 \\
  && BrPo &      6.007 &      5.676 &      8.103 &     10.912 &     11.949 \\
  \hline
  \multirow{3}{*}{1000}&
  \multirow{3}{*}{100}&
     CoSi &      \skip &      \skip &      \skip &      \skip &      \skip \\
  && KrNi &      \skip &      \skip &      \skip &      \skip &      \skip \\
  && BrPo &     29.651 &     27.393 &     41.917 &     59.072 &     70.150 \\
  \hline
\end{tabular}

  \caption{Benchmarks for the existing certified decision procedures.}
  \label{tbl:benchmarks}
\end{table}

The highly stochastic behaviour of the three algorithms makes this
data hard both to compute and to analyse: while the algorithms answer
in a reasonably short amount of time for a lot of pairs, there are a
few difficult pairs which require a lot of time (up to hours).
Therefore, we had to impose timeouts to perform these tests: a ``$>$''
symbol in Table~\ref{tbl:benchmarks} means that we only have a lower
bound for the corresponding cell. Also, since Coquand and Siles'
algorithm gives extremely bad performances for medium to large
expressions, we could not include timings for this algorithm in the
lower rows of this table.

The mean time is reported in the fourth column. Our implementation is
an order of magnitude faster than the other ones---even several orders
w.r.t.\ CoSi for non-trivial expressions.
%
However, this mean times are not representative of the actual
behaviour of the algorithms: they do not properly account for their
behaviour on the few difficult pairs which require a lot of time (both
because their weight is low since they are few, and because 500 pairs
are not enough to capture difficult pairs in a uniform way). This is
why we include the four remaining columns. For each of these columns,
say the one entitled ``90\%'', we computed the time which is
sufficient to solve at least 90\% of the pairs. In other words, the
column 50\% corresponds to the median times, the column 90\% to the
last deciles, 99\% to the last percentiles, and 100\% to the maximal
recorded times. For instance, with 20 nodes and 2 variables, 90\% of
pairs were solved within 0.152 seconds with KrNi; equivalently, 10\%
pairs required more than 0.152 seconds.

\begin{figure}
 \includegraphics[width=.8\linewidth]{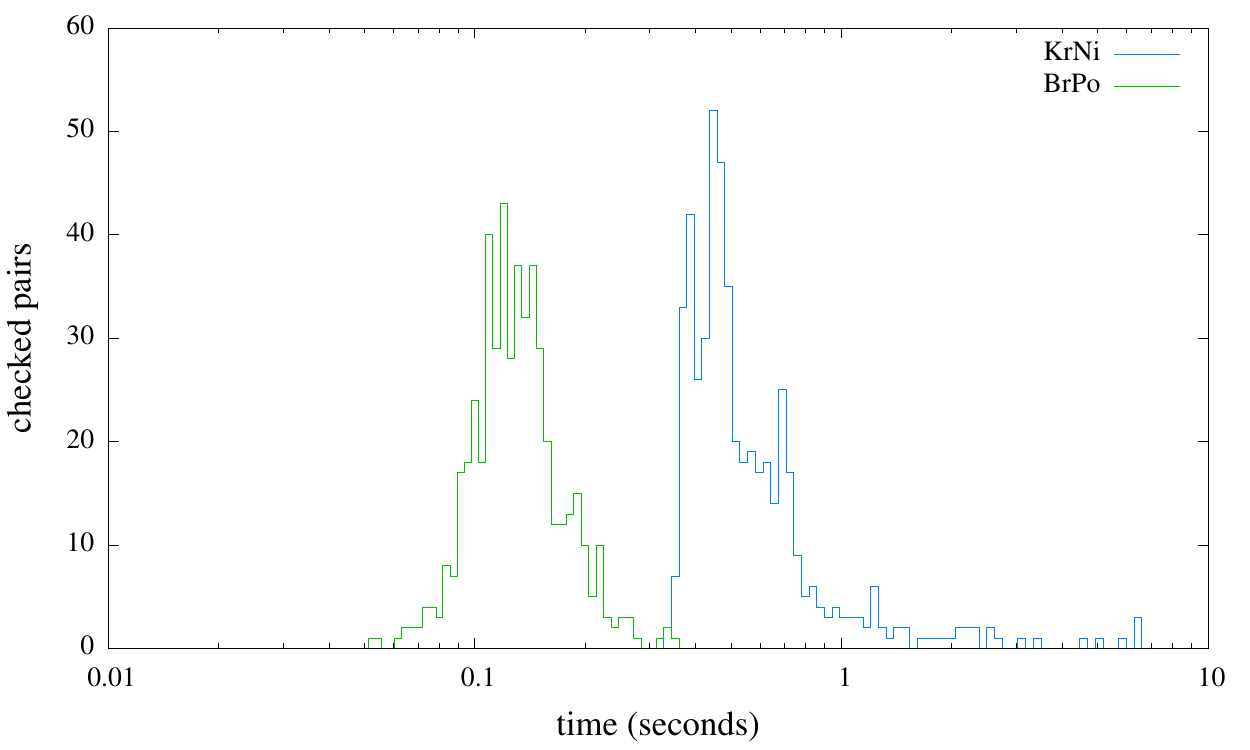}
 \caption{Distribution of the timings measured with Krauss and Nipkow'
   algorithm and ours (for the 500 pairs with 100 nodes and at most 10
   variables from Table~\ref{tbl:benchmarks}).}
  \label{fig:distribution}
\end{figure}

We also report in Fig.~\ref{fig:distribution} the distribution of the
timings we obtained for the pairs with 100 nodes and at most 10
variables, with KrNi and BrPo. 
These parameters correspond to the line in Table~\ref{tbl:benchmarks}
where the two algorithms are the closest in terms of performances; we
can however notice that while the median values are comparable, KrNi
suffers from a rather long trail: there is a difference of one order
of magnitude for the last percentile.

For larger expressions (500 to 1000 nodes), our tactic clearly
outperforms the two other ones, in terms of both mean time, median
time, and worst cases trail. In particular, our implementation seems
to be much more robust w.r.t.\ difficult pairs: in
Table~\ref{tbl:benchmarks}, the value of the last percentile is always
roughly equal to twice the median value, so that the mean is always
almost equal to the median.

The particular care we took to implement all steps of our procedure in
an efficient way could partially explain the observed performance gap;
however, our intuition is that this gap mainly comes from the
construction algorithm we use (by Ilie and
Yu~\cite{Ilie-yu-FollowAutomata}), which produces smaller
automata than the ones obtained with Brzozowski
derivatives~\cite{Brzozowski64}.

\section{Conclusions}
\label{sec:ccl}

We presented a correct and complete reflexive tactic for deciding
Kleene algebra equalities. This tactic belongs to a broader project
whose aim is to provide algebraic tools for working with binary
relations in Coq. The development is axiom-free, it can be downloaded
from~\cite{kleenecoq:web}. To our knowledge, this is the first
certified efficient implementation of these algorithms and their
integration as a generic tactic.

According to \texttt{coqwc}, the development consists of approximately
10.000 lines of Coq code, which distribute as follows and to which we
must add a 350 lines OCaml file for performing reification:

\medskip
\begin{center}
  \begin{tabular}{|l|r|r|r|}\hline
                       & specifications & proofs & comments \\\hline
    infrastructure     & 1959           & 1139   & 486      \\
    models             & 797            & 313    & 98       \\
    matrices           & 633            & 510    & 93       \\
    decision procedure & 1716           & 2353   & 261      \\\hline\hline
    total              & 5105           & 4315   & 938      \\\hline
  \end{tabular}
\end{center}
\medskip

The \emph{infrastructure} line corresponds to the basic infrastructure
files, the definition of the algebraic hierarchy using typeclasses,
and basic lemmas and tactics for monoids, semi-lattices, idempotent
semirings, and Kleene algebras. As expected, this part is rather
verbose. The \emph{models} line is for the definition of the various
models, including languages, binary relations, and regular expressions;
proofs are either trivial or fully automatised in this part.  The
\emph{matrices} line corresponds to all matrix constructions (up to
the fact that matrices form a Kleene algebra); proofs are eased by the
tactics we defined in the infrastructure but they are not fully
automatic: they follow standard paper proofs. The remaining line
corresponds to the decision procedure itself. As expected, this is
where the ratio proofs/specification is the largest: although we
exploit high-level tactics to perform case analyses, or \texttt{omega}
to reason about arithmetic, most proofs are non-trivial and have to be
rather explicit.

\subsection{Related works}
\label{ssec:related}

\subsubsection*{Algebraic tools for binary relations.}

The idea of reasoning about binary relations algebraically is
old~\cite{TG87,DBW97}. Among others~\cite{Kahl-ISAR,RALL}, Struth
applied this idea within an interactive theorem
prover~\cite{Struth-ChurchRosser-KleeneAlgebra}. He later turned to
automated first-order theorem provers (ATP): H\"ofner and him verified
facts about various relation
algebras~\cite{StruthHofner_KA,DBLP:conf/cade/HofnerS08} using
Prover9, a resolution/paramodulation based ATP. Our approaches are
quite different: we implemented a decision procedure for a decidable
theory, whereas their proposal consists in feeding a generic automated
prover with the axioms of some algebras, and to see how far the prover
can go by itself. As a consequence, their methodology applies directly
to a very wide class of goals and algebras, while we are restricted to
the equational theory of Kleene algebras. On the other hand, our
tactic always terminates, while Prover9 is unpredictable: even for
very simple goals, it can diverge, find a proof immediately, or find a
proof in a few minutes~\cite{DBLP:conf/cade/HofnerS08}.  Foster,
Struth, and Weber recently used Isabelle/HOL to formalise proofs about
relation algebras~\cite{FosterSW11}. While our long-term goals are
very close, our approaches and results are quite different, for the
same reasons as above: we focused on a single tactic to solve the
whole equational theory of Kleene algebra, while they use generic
automatic methods that are applicable to a much wider class of goals,
at the cost of requiring user-guidance if the goal is not simple
enough.

Narboux defined a set of Coq tactics for diagrammatic
proofs~\cite{narbouxPhD}. He works in the concrete setting of binary
relations, which makes it possible to represent more diagrams, but
does not scale to other models. The level of automation is rather low:
it basically reduces to a set of hints for the \coqe{auto} tactic.

\subsubsection*{Finite automata theory.}

The notion of strict star form (Sect.~\ref{ssec:epsilon}) was inspired
by the standard notion of \emph{star normal form}~\cite{klein93} and
the idea of \emph{star unavoidability}~\cite{Ilie-yu-FollowAutomata}.
To our knowledge, using this notion to get \eNFA{}s with acyclic
epsilon-transitions is a new idea.

At the time we started this project, Briais formalised decidability of
regular languages equality~\cite{grosquick} (but not Kozen's
initiality theorem). However, his approach is not computational, so
that even straightforward identities cannot be checked by letting Coq
compute.

The Isabelle/HOL tactic implemented by Nipkow and Krauss to decide
regular expressions equivalence~\cite{Krauss} is simpler than the one
we presented here, for several reasons. First, they implemented an
algorithm based on Brzozowski's
derivatives~\cite{Brzozowski64,Rutten98}, which is less involved than
ours, but also less efficient: the DFAs are produced directly from the
regular expressions, but they can be much
larger~\cite{Ilie-yu-FollowAutomata}. This certainly explains the
performance gaps we observed in Sect.~\ref{ssec:benchs}. Second, they
do not prove Kozen's initiality theorem: they prove correctness in the
model of regular languages and they use a nice mathematical trick to
reach the model of binary relations. As a consequence, their tactic
cannot be used with other models like matrices, $(min,+)$ algebras, or
weighted relations (graphs whose vertices are labelled by the elements
of an arbitrary Kleene algebra). Third, they do not formalise the
proof of completeness, or equivalently, the fact that the algorithm
always terminates (Isabelle/HOL computations do not need to terminate
so that they can use a ``while-option'' combinator). For all these
reasons, their development is much more concise than ours.

Coquand and Siles' recent implementation of the same algorithm than
Krauss and Nipkow in Coq~\cite{CoquandS11} is not efficient, and
cannot reliably be used for expressions with more than twenty nodes
(see Table~\ref{tbl:benchmarks}). A possible explanation could be that
they mix proofs and computations: this is known to be problematic
since proofs then have to be passed around along reductions, even with
\coqe{vm_compute}---the efficient Coq normalisation
function~\cite{GregoireL02}. Like Krauss and Nipkow, they do not
formalise Kozen's initiality theorem; they prove the completeness of
their algorithm, though.

\subsubsection*{Formalisation of algebraic hierarchies.}

The problem of formalising mathematical structures or algebraic
hierarchies in type theory is well-known and usually considered as
difficult~\cite{Barthe95, bt95, GeuversPWZ02, CoenT07,
  pack:math:struct}. Thanks to the recent addition of first-class
typeclasses~\cite{SozeauOury08}, we can use a very simple and naive
solution here, which gives us overloading for notations, lemmas, and
tactics, as well as modularity, sharing, and a basis for reification
(Sect.~\ref{sec:underlying}).

Since we started this project, Spitters and van der Weegen also
described how to use typeclasses to define an algebraic
hierarchy~\cite{math-classes}. Leaving apart the fact that we work
with typed structures, they follow the strategy we presented here (and
previously in~\cite{BraibantP10}); in particular, they use separate
classes for operations and laws, and they attach notations to class
projections. They actually use an even stronger discipline: each
operation comes with a class (e.g., our \coqe{Monoid_Ops} class
corresponds to their classes \code{SemiGroupOp} and
\code{MonoidUnit}).

We discussed two drawbacks of this approach in
Sect.~\ref{ssec:tc:details}, the most important one from our point of
view being the difficulty we had when trying to work with richer
structures. Indeed, the hierarchy we need for this work is really
small (it has depth three where the one from~\cite{pack:math:struct}
had depth ten at the time of writing), so that there are few instances
to declare for typeclass resolution. As a consequence, typeclass
resolution is efficient and the approach works out of the box. On the
contrary, our attempts to define richer structures were rather
frustrating. There are many more instances to declare (these include
all the inheritance relationships, all model constructions like
matrices, all the compatibility lemmas that give the ability to
rewrite using user-defined relations). Thus, typeclass resolution
becomes too slow to be used in practice---when we manage not to
introduce infinite loops, which also happens to be difficult.

Therefore, for rather large algebraic hierarchies, it is unclear to us
whether one should pursue with this simple approach, betting that
these problems can be resolved by improving the implementation of
typeclasses. Despite their apparent complexity, solutions like the
ones proposed in~\cite{pack:math:struct} might be less hazardous.

\subsection{Directions for future work}
\label{ssec:futurework}

We conclude with possible directions for future work.

\subsubsection*{Earlier failure checks.} 
Our algorithm for checking equivalence of DFAs returns whenever two
non-equivalent states are encountered. This makes the tactic faster in
case of failure, which is interesting when the tactic is used in a
``try'' block, where failures are expected to happen. We could
actually go one step further, by checking the equivalence on-the-fly,
during the determinisation phase. This means computing the DFAs lazily
and stopping as soon as a discrepancy is found. Doing so, we would
avoid the potentially expensive computation of the whole DFAs in case
of failure.
Although this approach is definitely more efficient than the current
one for the case of failures, it introduces some difficulties in the
correctness proof, which we did not complete.

\subsubsection*{A simpler proof of initiality.} 
Since we wanted to get a tactic for all models of Kleene algebras, we
had to formalise Kozen's initiality proof. With this goal in mind, the
derivative-based algorithm implemented by Nipkow and
Krauss~\cite{Krauss} is quite appealing for its simplicity. Moreover,
since the notion of derivative is purely syntactic, it is very well
suited to algebraic reasoning. However, rather surprisingly, we could
not find a way to replay Kozen's initiality proof with this algorithm.
We leave this question for future work.

\subsubsection*{KAT, Hoare logic.}
We plan to extend our decision procedure to deal with \emph{Kleene
  algebras with tests} (KAT), so as to provide automation to prove
correctness of programs in Hoare logic~\cite{kozen:hoare:kat:00}. A
first possibility would be to encode KAT expressions into
KA~\cite{KATCompletenessDecidability} and to use the current
tactic. This encoding being exponential in the number of predicate
variables, it is unclear whether this approach would be tractable. A
more involved approach would be to use the dedicated automata
construction presented in~\cite{KATComplexity}.

\subsubsection*{Richer algebras.}
Kleene algebras lack several important operations from binary
relations: intersection, converse, complement, residuals\dots{} We
plan to develop other tools for algebras dealing with these operators,
like \emph{Kleene algebras with converse}~\cite{CrvenkovicDE00},
\emph{residuated Kleene lattices}~\cite{Jipsen04}, or
\emph{allegories}~\cite{FreydScedrov90}. In particular, residuated
structures provide means of encoding properties like
well-foundedness~\cite{DBW97}, which are quite important for program
semantics. These structures are not known to be decidable; waiting for
new algorithms to be found, we can already build on our library to
implement various tools for working with these structures in the Coq
proof assistant.

\bigskip\noindent \textbf{Acknowledgements.}  We warmly thank Guilhem
Moulin, Assia Mahboubi, Matthieu Sozeau, Bruno Barras, and Hugo
Herbelin for highly stimulating discussions. We are also grateful to
the anonymous referees of the first Coq workshop in 2009 and ITP in
2010, whose remarks helped us to improve both the library and its
description.

\bibliographystyle{plain}
\bibliography{bib}

\end{document}